%% file: Paper_DouglasQuaid_Good_Archivx_strip.tex
\documentclass{article}

\usepackage{arxiv}
\usepackage[utf8]{inputenc}
\usepackage[T1]{fontenc}
\usepackage{hyperref}
\usepackage{url}
\usepackage{booktabs}
\usepackage{amsfonts}
\usepackage{nicefrac}
\usepackage{microtype}
\usepackage{xcolor}
\usepackage{graphicx}
\usepackage{subcaption}
\usepackage{enumitem}
\usepackage{wrapfig}
\usepackage{listings}
\usepackage{rotating}
\usepackage[autostyle=true]{csquotes}

\usepackage[xindy]{glossaries}
\makeglossaries
\setglossarystyle{listgroup}

\loadglsentries{glossaryEN}

\DeclareUnicodeCharacter{FB01}{fi}

\title{Douglas-Quaid - Open Source Image Matching Library}

\author{
  Vincent~Falconieri\\
  CIRCL\\
  Luxembourg\\
}

\begin{document}

\maketitle

\begin{abstract}
Security analysts need to classify, search and correlate numerous images. Automatic classification tools improve the efficiency of such tasks. However, no open-source and turnkey library was found able to reach this goal. The present paper introduces an Open-Source modular library for the specific cases of visual correlation and Image Matching named \textit{Douglas-Quaid}.
The design of the library, chosen tradeoffs, encountered challenges, envisioned solutions as well as quality and speed results are presented in this paper.
We also explore researches directions and future potential developments of the library. Our claim is that even partial automation of screenshots classification would reduce the burden on security teams and that \textit{Douglas-Quaid} is a step forward in this direction.
\end{abstract}


\section{Introduction}
CERTs - as CIRCL - and security teams collect and process content such as images (at large from photos, screenshots of websites or screenshots of sandboxes). Datasets become larger - e.g. on average 10000 screenshots of onion domains websites are scrapped each day in AIL, an analysis tool of information leak - and analysts need to classify, search and correlate through all the images.

Automatic tools can help them in this task. Less research about image matching and image classification seems to have been conducted  exclusively on websites screenshots. \cite{sampatCNNTaskClassification}\cite{aburrousPredictingPhishingWebsites2010a}\cite{chenFightingPhishingDiscriminative2009}. However, a classification of this kind of pictures needs to be addressed.

One of our long-term goal was to build a generic library and services which can easily be integrated in \textit{Threat Intelligence tools} tools such as 
\textbf{AIL}\footnote{Analysis Information Leak framework - \href{https://github.com/CIRCL/AIL-framework}{github.com/CIRCL/AIL-framework}}\cite{mokaddemAILDesignImplementation2018} and 
\textbf{MISP}\footnote{Malware Information Sharing Platform - \href{https://github.com/MISP/MISP}{github.com/MISP/MISP}}\cite{wagnerMISPDesignImplementation2016}.

\glsenablehyper\gls{MISP} is an open source software solution tool developed at CIRCL for collecting, storing, distributing and sharing cyber security indicators and threats about cyber-security incidents analysis. \\
\glsenablehyper\gls{AIL} is an open source modular framework developed at CIRCL to analyze potential information leaks from unstructured data sources or streams. It can be used, for example, for data leak prevention.

A first step of this long-term goal completion had been reached with the development of an evaluation framework, provided as \textit{Carl-Hauser}\footnote{\href{https://github.com/CIRCL/carl-hauser}{github.com/CIRCL/carl-hauser}} and an open-source library itself, provided as \textit{Douglas-Quaid}\footnote{\href{https://github.com/CIRCL/douglas-quaid}{github.com/CIRCL/douglas-quaid}}. This paper presents the design and the current performances of \textit{Douglas-Quaid} library.

Image-matching algorithms benchmarks already exist \cite{gaillardLargeScaleReverse2017}\cite{zaunerImplementationBenchmarkingPerceptual2010}\cite{bianImageMatchingApplicationoriented2017} and were highly informative. They were used to choose on which algorithms to focus on. A quick-lookup mechanism for correlation was necessary and took part in this library.

\subsection{Problem Statement}
No open-source tool provides an easy high level interface to correlate pictures, even without conditions on technology used.

Ideally, establishing links or correlation between pictures could be fully automated, to assist analysts in their duty. Even partial automation would reduce the burden of this task on security teams. \cite{cevikalpLargescaleImageRetrieval2018} states the \glsenablehyper\gls{Image-Retrieval} problem as "Given a query image, finding and representing (in an ordered manner) the images depicting the same scene or objects in large unordered image collections". 

\textbf{The main contribution of this paper is a free and open-source library for \glsenablehyper\gls{Image-Matching} with a high level API.}

This paper also presents the design of this system, used algorithms, possible and expected extensions, as well as performance results.

\section{Methodology and Context}

\subsection{Context}

\begin{figure}[h!]
\centering 
\includegraphics[width=0.9\textwidth]{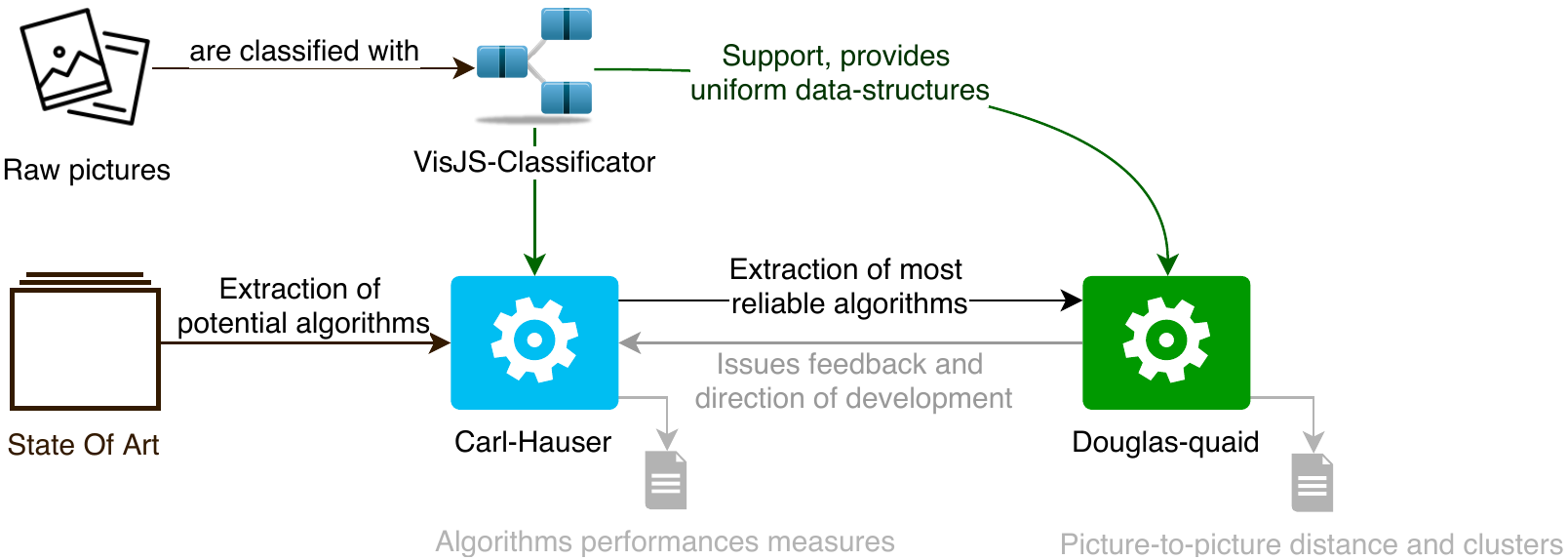} 
\caption{Overview of projects}
\label{projectsoverview}
\end{figure}

\glsenablehyper\gls{Douglas-Quaid} had been built among other projects, related to the same problem and influencing its design choices; as shown in Figure \ref{projectsoverview}.

We completed a State of Art about Image Matching algorithms available at \href{https://www.github.com/Vincent-CIRCL/carl-hauser/blob/master/SOTA/SOTA.pdf}{SOTA.pdf}. This State of Art allowed us to build a framework named \glsenablehyper\gls{Carl-Hauser}\cite{190803449CarlHauser}, which evaluate literature's algorithms with the most potential for our use-case.

To classify the datasets of pictures needed to evaluate algorithms performances, we built \glsenablehyper\gls{VisJS-Classificator}\cite{falconieriVisJSClassificatorManualVisual2019} to visually classify pictures under a graph data-structure as well as a clustered data-structure, at the same time. More information about the dataset is provided in Section \ref{datasetinfo}.

\glsenablehyper\gls{Douglas-Quaid} was then built as a performance-oriented implementation of  \glsenablehyper\gls{Carl-Hauser}'s best \glsenablehyper\gls{Image-Matching} algorithms. To date, \glsenablehyper\gls{Carl-Hauser} is still more advanced than \glsenablehyper\gls{Douglas-Quaid} in terms of algorithms complexity. However, the gap may shrink in near future. \glsenablehyper\gls{Douglas-Quaid} is much faster than \glsenablehyper\gls{Carl-Hauser}, as he is performance-oriented, rather than exploration-oriented.

\subsection{Datasets}
\label{datasetinfo}
Tests, performances and speed evaluation were conducted using real sets of pictures. These datasets were extracted from CIRCL's tools, such as \glsenablehyper\gls{AIL} and \glsenablehyper\gls{URLAbuse}. 
Two main datasets were used : \href{https://www.circl.lu/opendata/circl-ail-dataset-01/}{\textit{circl-ail-dataset-01}} of 470+ pictures and \href{https://www.circl.lu/opendata/circl-phishing-dataset-01/}{\textit{circl-phishing-dataset-01}} of 37000+ pictures.

Phishing datasets and AIL datasets are available for research purposes at \href{https://www.circl.lu/opendata/datasets/circl-phishing-dataset-01/}{\textit{circl.lu/opendata/datasets/circl-phishing-dataset-01}} and \href{https://www.circl.lu/opendata/datasets/circl-ail-dataset-01/}{\textit{circl.lu/opendata/datasets/circl-ail-dataset-01}}.
More details about these datasets in \cite{falconieriOpenDatasetPhishing2019}.

\begin{figure}[h!]
  \centering
  \begin{subfigure}[b]{0.49\linewidth}
    \includegraphics[width=\linewidth]{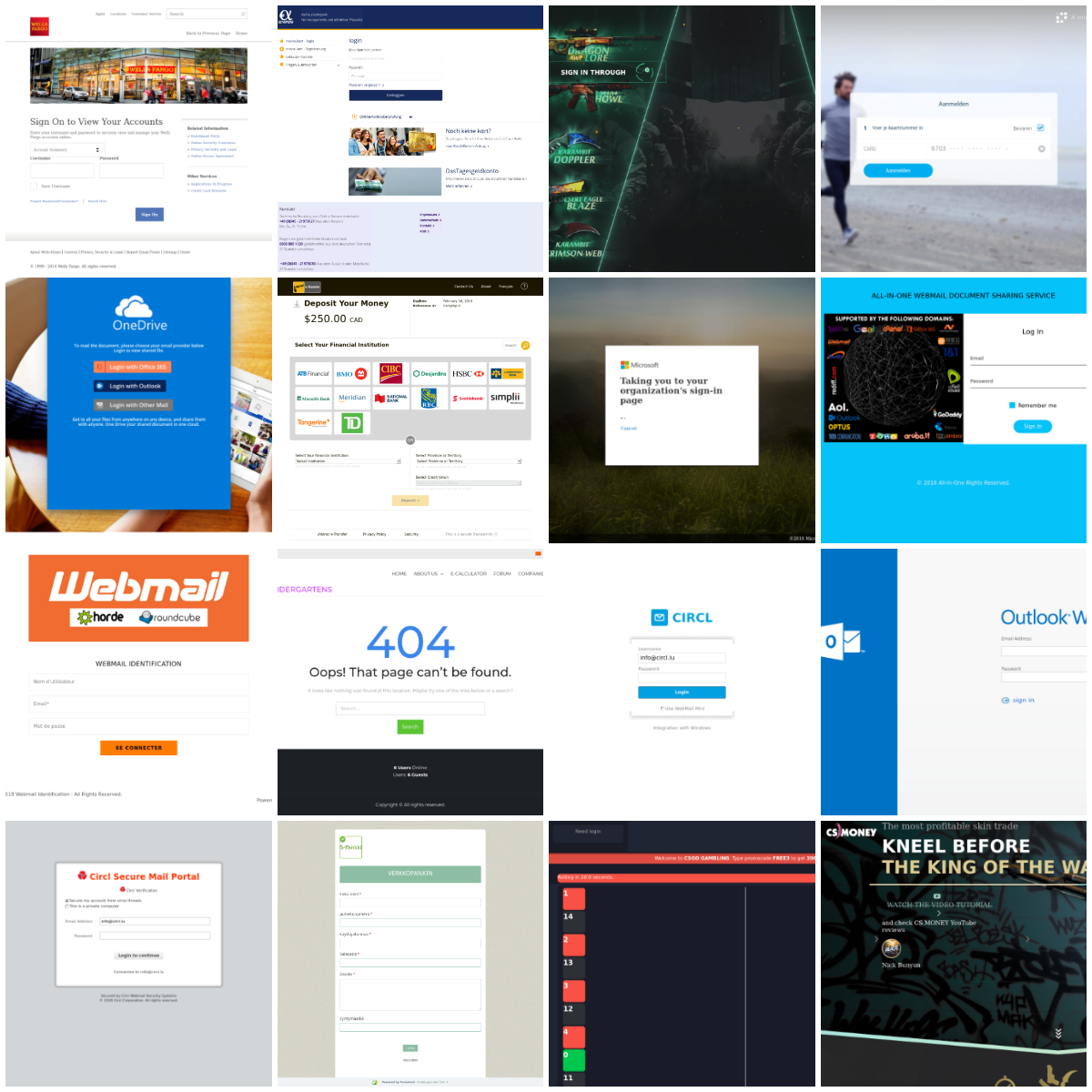}
    \caption{Phishing dataset overview (470+ pictures)}
  \end{subfigure}
  \begin{subfigure}[b]{0.49\linewidth}
    \includegraphics[width=\linewidth]{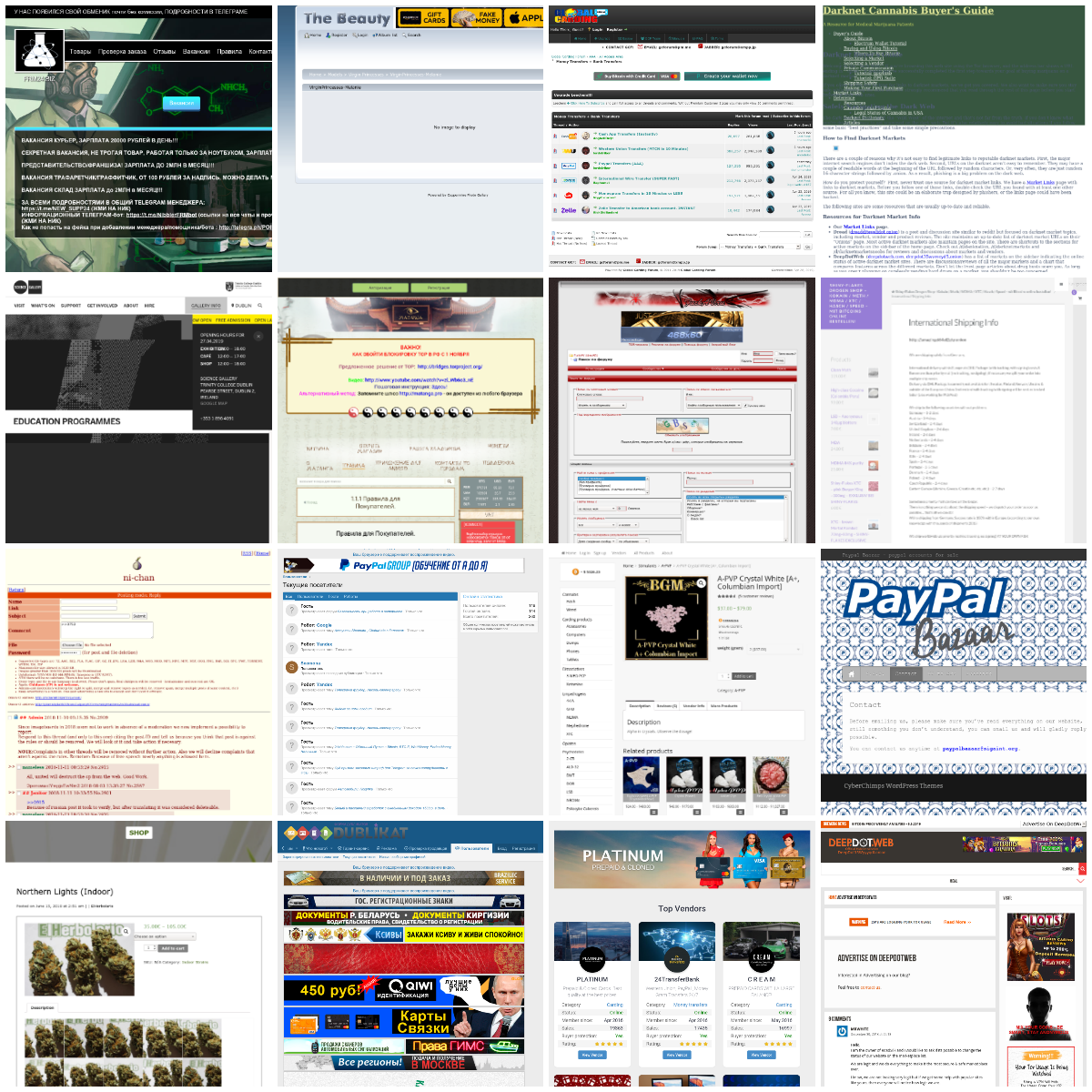}
    \caption{AIL dataset overview (37000+ pictures)}
  \end{subfigure}
  \caption{Dataset's samples}
  \label{fig:datasetsextract}
\end{figure}

\section{Design}

\glsenablehyper\gls{Carl-Hauser} framework allowed us to know the pros and cons of each algorithm and approach. It acted as a precursory of \glsenablehyper\gls{Douglas-Quaid} library. 

\begin{wrapfigure}{r}{0.3\textwidth}
\includegraphics[width=0.3\textwidth]{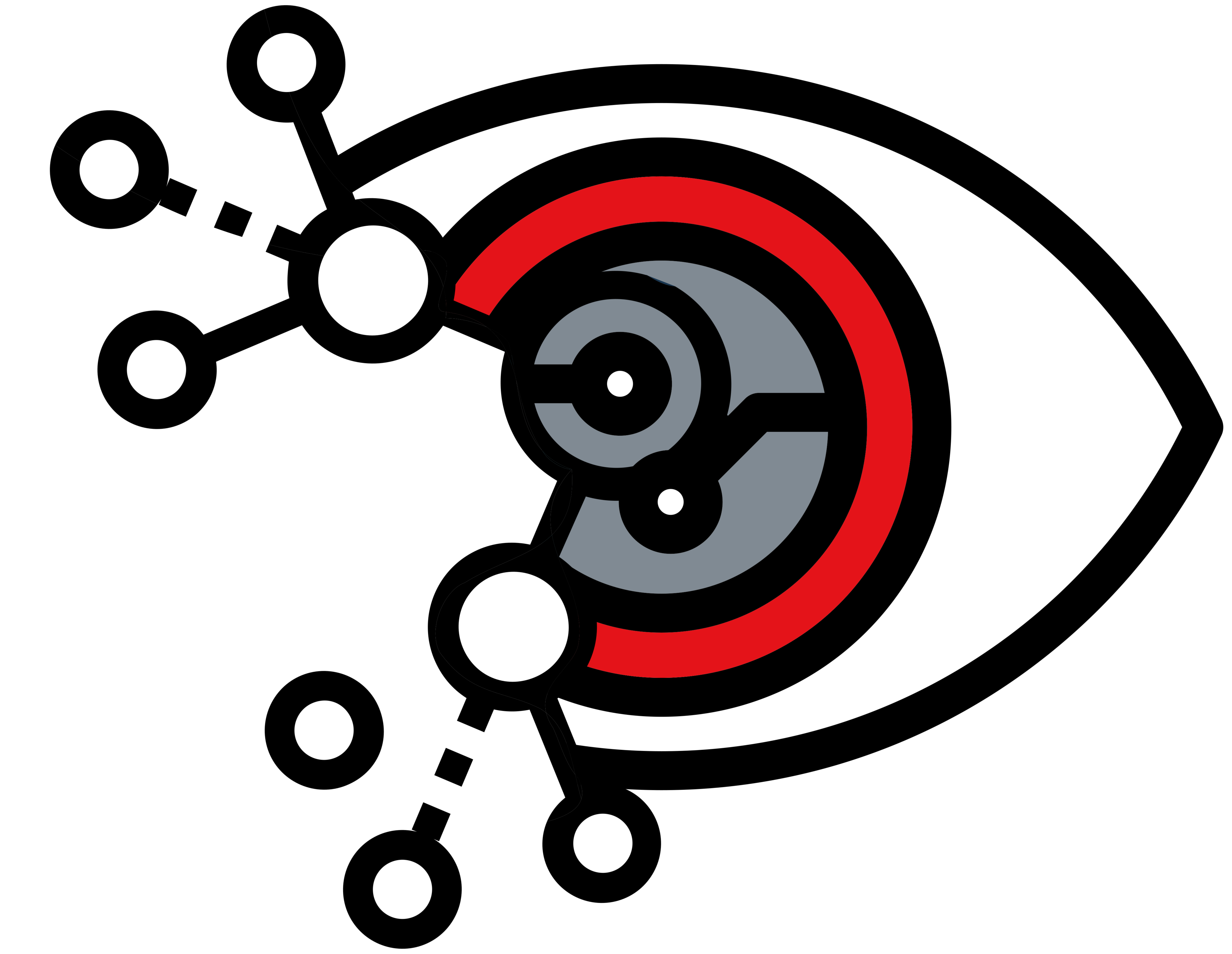}
\caption{Douglas-Quaid logo} 
\end{wrapfigure}

\subsection{Hypothesis, constraints and goals}
We set main assumptions and goals as well as standard software design expectations (maintainability, quality of code, documentation ...). These constraints were : 

\begin{itemize}[noitemsep]
\item No Open-Source Image-Matching library with High-level API were found. API calls should be kept to minimum and as simple as "\textit{store a picture}", "\textit{request similar pictures}", etc.
\item From Carl-Hauser, we found out that chosen algorithms are complementary. An approach with a unique or a hardcoded set of algorithms would be restrictive. Core algorithms should be modular.
\item Nature of pictures varies. Even if the goal is to match screenshots, nature of these screenshots varies : phishing websites, infected virtual-machines, onion domain websites ... Therefore, the behavior of the library needs to be altered easily, an easily specialized for one use.
\item Technologies should easily interact with other Open-Source tools developed at CIRCL, if possible.
\item The library must scale up, especially for read/request accesses. A request should be answered quite quickly, to be usable to automate processes as well as used by analysts.
\end{itemize}

\subsection{Software Overview}

Like \glsenablehyper\gls{Carl-Hauser}, \glsenablehyper\gls{Douglas-Quaid} has therefore been implemented from scratch. Similar good practices were followed \cite{RepositoryStructurePython}\cite{linGoodLoggingPractice}. 

\glsenablehyper\gls{Douglas-Quaid} can be considered as a less algorithmically complex version of \glsenablehyper\gls{Carl-Hauser}. \glsenablehyper\gls{Carl-Hauser} is a research platform that evaluates very heterogeneous treatments, while \glsenablehyper\gls{Douglas-Quaid} only implements the most robust versions of \glsenablehyper\gls{Carl-Hauser} algorithms. 

Core database, processing and results storage can be distributed on different machines, as \glsenablehyper\gls{Douglas-Quaid} is built as a client and multi-tier server architecture. The REST API constitutes the main mean of interaction with \glsenablehyper\gls{Douglas-Quaid}'s server. Entirely Local  dynamic server instantiation is available, for instance to automatically evaluate optimal parameters.

the storage uses \glsenablehyper\gls{Redis}, which is an in-memory database. Only set of feature - not images - are stored in memory (about 45KB/image added). The reactivity of \glsenablehyper\gls{Redis} was chosen at the expense of the "infinite" scalability of the library. Three Redis databases are parts of the server: a test database (optimization measures, unit tests, etc.), a cache database (used to queue initial or intermediate requests, etc.) and a storage database (stores the main data structure).

The treatments are monothreaded, per worker. A variable number of workers can be started, potentially bounded by Redis to handle their requests. Each worker has a specific task, that he performs in a loop. Queuing structures are used to assign tasks to workers. Hot-shutting down and hot-adding of workers can be performed as needed. For example, during a data addition phase (write intensive), many "add" workers can be started simultaneously. Access to the add queue is protected to ensure that jobs are handled only once. A double addition would in any case not be an issue for the main data structure.

The query results are stored in the cache database, waiting for the client who ordered it to pick it up. Most intermediate data have an expiration time, avoiding a constant and uncontrolled increase in the memory used in case of issues with workers.

The API is simplistic and the complexity moved to the server side. Each implemented algorithm can be enabled or disabled and has specific parameters. The choice of these parameters is difficult, but is solved by a calibration algorithm. Process followed by this calibration algorithm is presented by Figure \ref{calibrationprocessoverview}. 

The library therefore only needs:
\begin{itemize}[noitemsep]
\item a subset of the production dataset (for example, 20 to 40 images sampled from a complete dataset);
\item a "ground truth" file to define the ideal result expected;
\item the target rates of false positive/true positive, false negative/true negative (at least two of them).
\end{itemize}

The calibration algorithm will seek to optimize the internal parameters of \glsenablehyper\gls{Douglas-Quaid} with these inputs. The calibration algorithm is described in Section \ref{selfcalibration}. The main advantage is therefore the absence of obscure parameters (example: a threshold between 0 and 1) to be provided without knowledge of the internal mechanisms of the library.

The main data structure is described in Section \ref{maindatastructure} and is not required to understand the general architecture of the library. This data structure simply makes it possible to obtain an expected sub-linear complexity in the number of images stored in the database, at best equal to the number of existing clusters. At best, the number of clusters and so the search complexity will be of $O(\sqrt{nb pictures in database})$.

If a parameter change takes place during the production phase, all data must be reloaded into the library. This operation is not implemented by a method to "recalculate" data structures to avoid adding an additional layer of complexity, which could generate artifacts. In order to improve maintainability, a change in core parameters must be followed by a restart (and flush) of the server.

The technical documentation of \glsenablehyper\gls{Douglas-Quaid} is available at \url{https://github.com/Vincent-CIRCL/douglas-quaid/blob/master/docs/code_doc/core_doc.pdf}

\clearpage
\pagebreak
\subsection{Software design}

\begin{figure}[h!]
\centering 
\includegraphics[width=\textwidth]{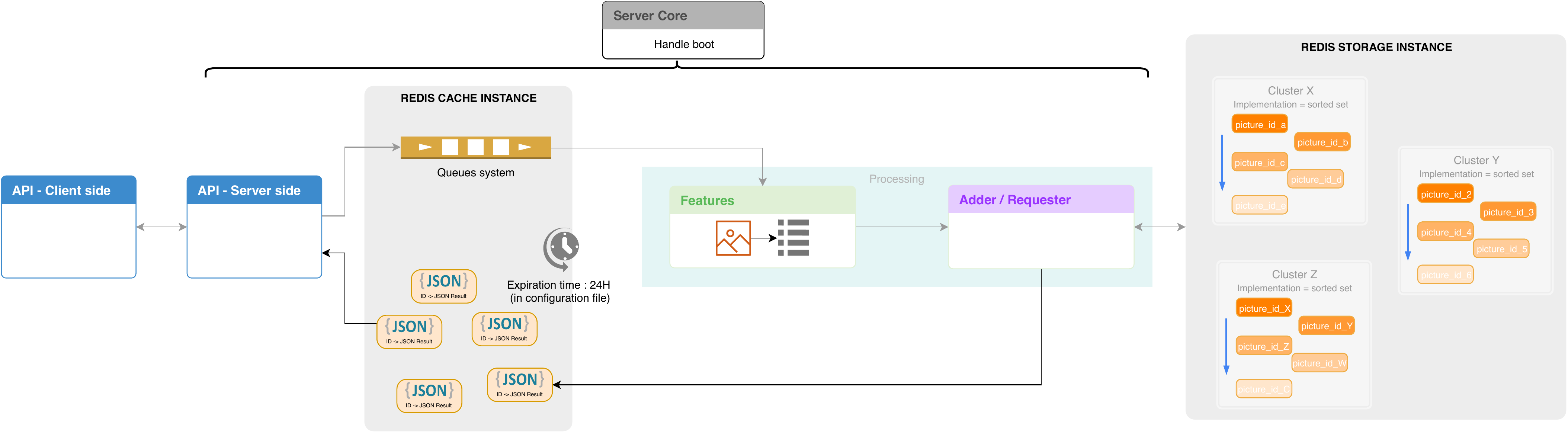} 
\caption{Overview of Douglas-Quaid Library software architecture}
\label{libarchitectureoverview}
\end{figure}

The software architecture of the library is shown in a simplified way in Figure \ref{libarchitectureoverview}.

\subsubsection{Client side}

On the client side of the application we find a few main blocks : 
\begin{itemize}[noitemsep]
\item \textbf{API} (Application programming interface), with a Simple API and an Extended API. 
	\begin{itemize}[noitemsep]
	\item The Simple API presents the essential elements of the API, low level, to talk with the server. It gives small and essentials methods to exchange information with the server-side API. 
	\item The Extended API presents higher level methods to interact with the server-side API, which performs batch operation on folder for example.
	\end{itemize}
\item \textbf{CLI} (Command-line interface) for client side. It's a python script that can be launched with arguments (-h to get the help for details) to perform each action without need of a full custom python script.
\item \textbf{Similarity Graph extractor} and \textbf{Storage graph extractor}. Two denomination are used in the library : similarity and storage graph. Both are different and should not be mistaken. More information provided about these graph in \ref{graphdetails}
\item \textbf{Example} of client side API usage.
\end{itemize}

\subsubsection{Server side}

On the server side we find a few application blocks. It is composed of some classes to handle API calls, some to handle how to interact with the database, some to compute and extract features out of pictures, some to extract distances out of features, some to handle workers and processes, some to handle export and import ...

\begin{itemize}[noitemsep]
\item\textbf{ Server Side API} : This set of classes handle the reception and the processing of user's requests. It answers to API call, defines and instantiates server side endpoints for each API call, handles SSL. Can interact and call relevant functions elsewhere on the server-side.
\item \textbf{A Queuing system} : the API can queue a job that will be performed by a backend worker. It uses a Queuing system within a Redis database. These workers handles the transformation of a picture into a set of features, and handle the comparison between stored and requested set of features. The queuing system is also used to store results to be fetched by clients.
\item \textbf{A Storage database} with specifics data structures, to reach a square root search complexity in number of pictures stored. 
\item \textbf{A server's minimal core of Singletons} handles launch, verification and stop of databases, workers and processes. These launchers propagates inputted configuration files on launch, to alter workers behavior.
\end{itemize}

The entry points to launch \glsenablehyper\gls{Douglas-Quaid} server are : 
\begin{itemize}[noitemsep]
\item \textbf{Standard Server Launch} : ready to answer request in production environment
\item \textbf{Calibrator} : Behavior explained in Section \ref{selfcalibration}
\item \textbf{Scalability Evaluator} : Which is a set of classes used to evaluate the scalability of the application. Add and request pictures, measuring response time. It stores results in JSON and save it as a graph. It can evaluate the scalability on the database with default configuration or with different thresholds for cluster creation, which are specified in the configuration file.
\item \textbf{Test instance launcher} : Provides an unified way to create a test instance of the server. Point the storage and cache database (and sockets) to a test database, that won't modify production data. Allows to overwrite default configuration files. Operates with an overwritten version of database Default Configuration, modifying path of scripts and sockets. Create a running instance of douglas-quaid, all linked on a unique test database. Modify the behavior of the core launcher handler, to use only one database.
\end{itemize}

\subsubsection{Graphs}
\label{graphdetails}
Two denomination are used in the library : similarity and storage graphs. Both are different and should not be mistaken: 
\begin{itemize}[noitemsep]
\item \textbf{Storage graphs} : Graphs and clusters as it is stored in Redis. (Datastructure to improve requests performance). The storage graph refers to the set of clusters and their content, used in the database, to store pictures. It is a visualization of "how is the database" and can be analyzed to see if the database will be efficient or not. Classes on client side with this name are able to extract this graph from the server. Details about storage graph are given in Section \ref{maindatastructure}.
\item \textbf{Similarity graphs} : Graphs computed from many requests, that show which picture is close to which picture. This is not the way it is stored in the database, but a condensed view of all requests that can be made on the database. The similarity graph refers to the distance between pictures, and the graph its constitutes, if we extract each picture of the database and its nearest neighbor, as a client. It is a visualization of "what the database says" and can be analyzed to see if the library output is correct or not. Classes on client side with this name are able to extract this graph from the server.
\end{itemize}

\subsection{Algorithms}

\begin{wrapfigure}{r}{0.4\textwidth}
\includegraphics[width=0.4\textwidth]{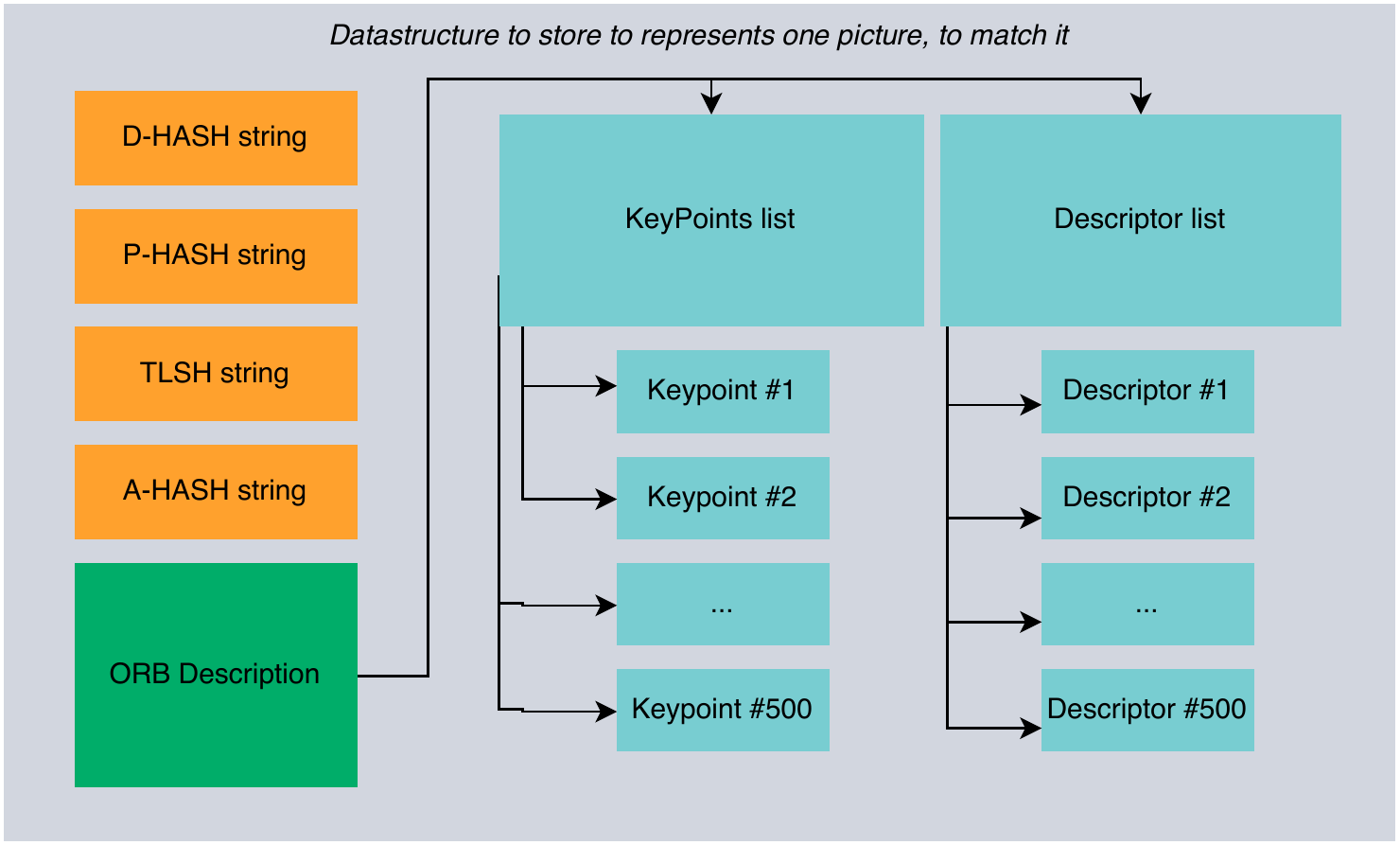} 
\caption{Representation of one picture}
\end{wrapfigure}

Few Image matching libraries were tested, including : 
\begin{itemize}[noitemsep]
\item \textbf{ImageHash}\footnote{\href{https://github.com/JohannesBuchner/imagehash}{github.com/JohannesBuchner/imagehash}} which includes a wide list of fuzzy hash algorithm such as AHash, DHash, PHash, WHash, ... The purpose of these algorithms is to map input files (here, images) into a limited hash space, such that two "similar" images have similar hashes. Their objective is therefore distinct from traditional hash algorithms, which seek to maximize the difference between hashes for even a minimal difference (ultimately, 1 bit) in the input file.
\item \textbf{TLSH}\footnote{\href{https://github.com/trendmicro/tlsh}{github.com/trendmicro/tlsh}}\cite{oliverTLSHLocalitySensitive2013} is also a fuzzy hashing algorithm, available in its own library.
\item \textbf{OpenCV}\footnote{\href{https://github.com/opencv/opencv}{github.com/opencv/opencv}} is an open source computer vision and machine learning software library, which provides a common infrastructure for computer vision applications. Relevant algorithms are available within it, such as SIFT \cite{loweDistinctiveImageFeatures2004}\cite{oteroAnatomySIFTMethod2014} (Scale-invariant Feature Transform), SURF \cite{baySURFSpeededRobust2006} (Speeded Up Robust Features), ORB \cite{rubleeORBEfficientAlternative2011} (Oriented FAST and Rotated BRIEF), ... These algorithms use key point descriptors, a way to condense distributed and locally relevant information, into a vector or set of vectors. These vectors can then be compared.
 However, we focused on open-source implementations and avoided patented implementation. SIFT and SURF were therefore outside of the scope. Thanks to a large performance overview \cite{tareenComparativeAnalysisSIFT2018} we chose to primarily focus on ORB.
\end{itemize} 

Usually, you may need to see (through \glsenablehyper\gls{VisJS-Classificator} for example, as the export format is compatible) what the Similarity Graph looks like. Except for debugging, Storage Graph does not need to be extracted in production settings.

\clearpage
\pagebreak
\section{Challenges}

\subsection{Main data-structure}
\label{maindatastructure}

The main data-structure is composed of \textbf{cluster of pictures}. Each cluster is the representation of a graph, with edges representing distances between pictures. As the exact graph is not needed, we don't store it under this format. Each cluster of picture is represented as a sorted set of pictures. Semantically we are working with graphs, while we are practically working with sets. 

The higher in the set a picture is, the most representative the picture is. Pictures in sorted sets are sorted by their representativeness of the whole set of pictures. Therefore, we have a sorted list of representative pictures of each cluster. Strictly, the representativeness is computed as a centrality measure of a given picture in the similarity graph. 

When a picture is added in the database or a request is made, algorithms perform a comparison of the picture with all most central (representative) picture(s) of each cluster. Therefore, instead of making a comparison of one pictures to all pictures of the database ($O(\textit{nb picture in db})$), it compares the picture with all clusters ($O(\textit{nb clusters} * \textit{nb most representative picture per cluster to test (PARAMETER)})$). 

The picture will be matched to a cluster (strictly, the representative picture of the cluster) or not. 

If the picture match a representative picture of a cluster, we compare the request picture with all pictures of this cluster. Therefore, instead of running $O(\textit{nb picture in db})$ comparisons, we run $O(\textit{nb clusters}+\textit{nb matched clusters}*\textit{nb pictures in each looked-up cluster})$. This improves performances by decreasing number of performed comparisons. At best, the number of comparisons will be $O(\sqrt{\textit{number of pictures stored}})$

Clusters are composed of somewhat-similar pictures. Tricky thing is the algorithm which says "\textit{This picture is too distant, it must be in a new cluster}" or "\textit{This picture is close enough and is in the same cluster}". Mainly based on a threshold, it impacts the performance of the library (time-wise) and less/not on the quality of match.  

\begin{figure}[h!]
\centering 
\includegraphics[width=\textwidth]{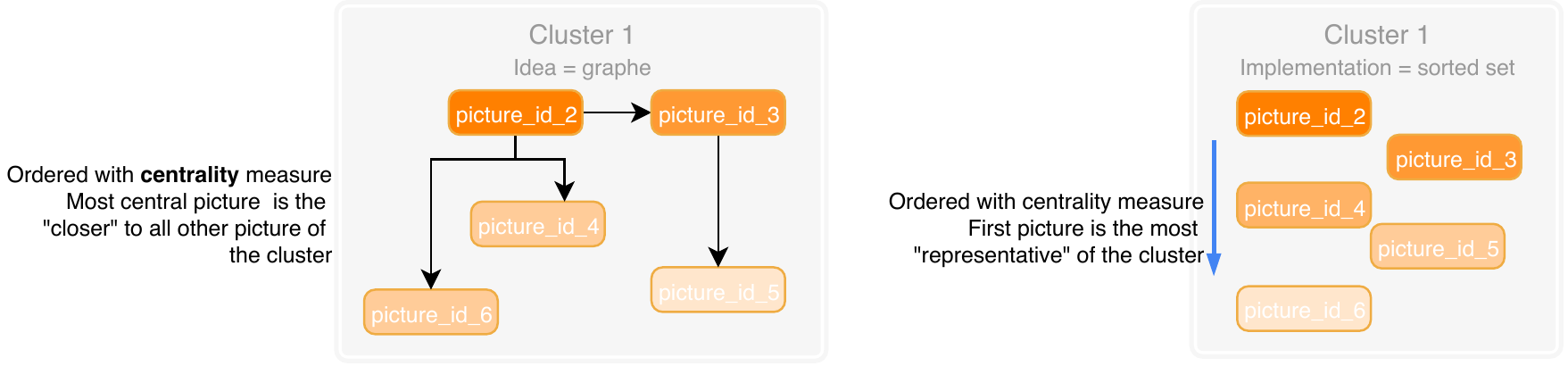} 
\caption{Conceptual view versus Implementation view}
\end{figure}

\subsection{Distances}
\label{distancesheterogeinity}

\begin{wrapfigure}{r}{0.5\textwidth}
\includegraphics[width=0.5\textwidth]{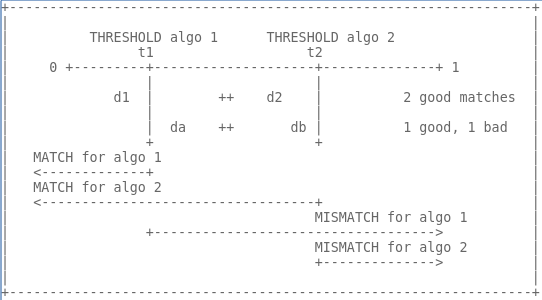} 
\caption{Distance homogeneity problem}
\label{distancehomonerinitygraph}
\end{wrapfigure}

Distances are needed but insufficient to represent link between pictures. Heterogeneity of algorithm's output is a problem, as the following will show. 

Please note that distances are considered here as normalized euclidean distance, which means that a 0 distance is equivalent to "two totally similar pictures", and a 1 distance is equivalent to "two totally different pictures". Let's consider an edge case.

If you consider two algorithms:
\begin{itemize}[noitemsep]
\item For algorithm A, a match is meaningful if the returned distance is below $0.25$. It is a mismatch if the distance if upper than $0.25$. Noted as $t1$ on Figure \ref{distancehomonerinitygraph}.
\item For algorithm B, a match is meaningful if the returned distance is below $0.85$. It is a mismatch if the distance if upper than $0.85$. Noted as $t2$ on Figure \ref{distancehomonerinitygraph}.
\end{itemize}

Consider two comparison : 
\begin{itemize}[noitemsep]
\item Algorithm A gives a distance of $0.2$ (match for itself), algorithm B gives a distance of $0.8$ (match for itself).  Noted as $d1$ and $d2$ on Figure \ref{distancehomonerinitygraph}.
\item Algorithm A gives a distance of $0.3$ (mismatch for itself), algorithm B gives a distance of $0.7$ (match for itself). Noted as $da$ and $db$ on Figure \ref{distancehomonerinitygraph}.
\end{itemize}

Min, max, mean ($0.5$) are equal in both cases. However, in first case, all algorithms agreed that both pictures are "a match". In second case, one algorithms considered it as a "mismatch". We do not even consider "gray zone" in this example, but only strict thresholds.  
This problem exists because each algorithm has its own "range of values" in which a match is a meaningful match. Normalization could have been a solution, but we have chosen a more human-readable solution : decisions.     

\subsection{Decisions}
\label{decisionsnecessity}

We choose to work with human-readable decisions in addition to distances. Decisions are \textbf{not} a simple threshold over distance. 

Each algorithm computes a distance between pictures. Each algorithm has it's own way to compute this distance. All algorithms output a value within range '[0-1]'. \\
For example : \textit{A-HASH outputs a distance of $0.4$ between picture A and B}

Each algorithm can then computes a decision. Using thresholds on its own distance output, it gives a YES/MAYBE/NO decision. \\
Continuing example : \textit{A-HASH outputs a decision 'YES (it matches)' between picture A and B, because distance between picture A and B is $0.4$, which is less than the '\textit{yes-to-maybe threshold}' of A-HASH. }

Decisions may help to decide which other algorithms to launch.     
That's why a "YES" or "MAYBE" decision can come after a "NO" decision in a distance-sorted list.

A word about precision: given the precision of enabled algorithm ($1/400$th precision for TLSH, $1/500$th precision for ORB), we can consider that any distance more precise than $1/250$th ($0.004$) precision would not be relevant. 

\subsection{Merging distances and decisions}
\label{mergingdistanceanddecisions}
When comparing two pictures, enabled algorithms are called. Each provides a distance and a decision. An additional merging algorithm do merge these distances and do merge these decisions.  

Distances can be merged with distinct approaches : 
\begin{itemize}[noitemsep]
\item Max of all distances;
\item Mean of all distances;
\item Min of all distances;
\item Harmonic mean;
\item Weighted mean of all distances, etc.  
\end{itemize}

\begin{figure}[h!]
\centering 
\begin{subfigure}[b]{0.33\textwidth} \centering 
\includegraphics[width=\textwidth]{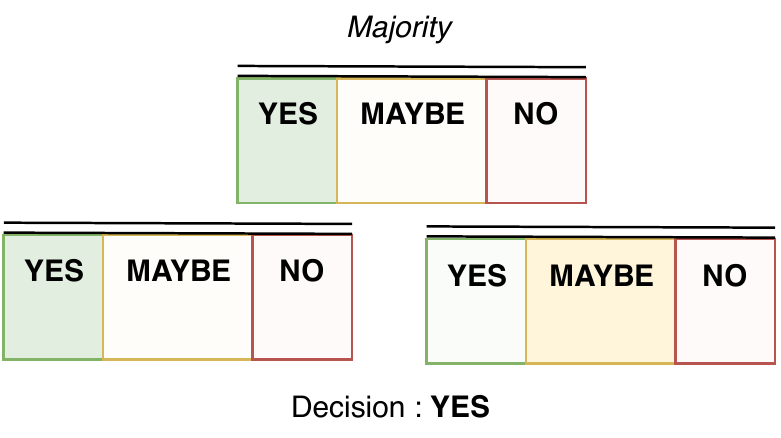} 
\caption{Majority method} \end{subfigure}
\begin{subfigure}[b]{0.33\textwidth} \centering 
\includegraphics[width=\textwidth]{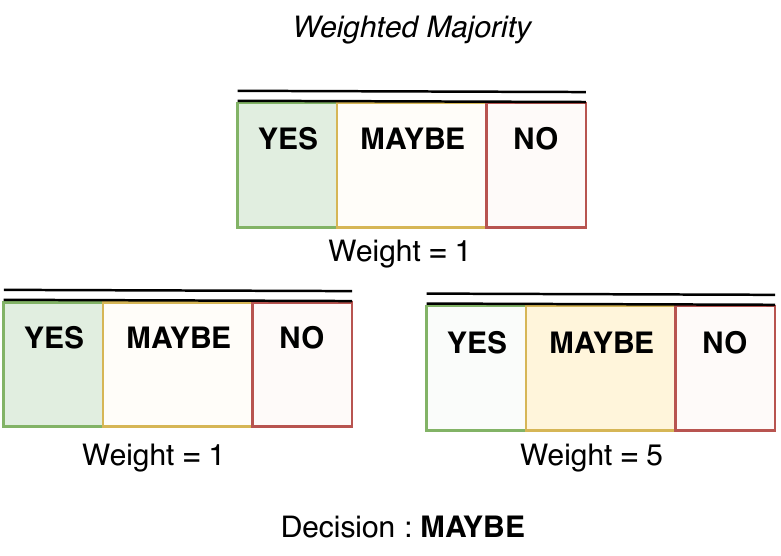} 
\caption{Weighted Majority method} \end{subfigure}
\begin{subfigure}[b]{0.32\textwidth} \centering 
\includegraphics[width=\textwidth]{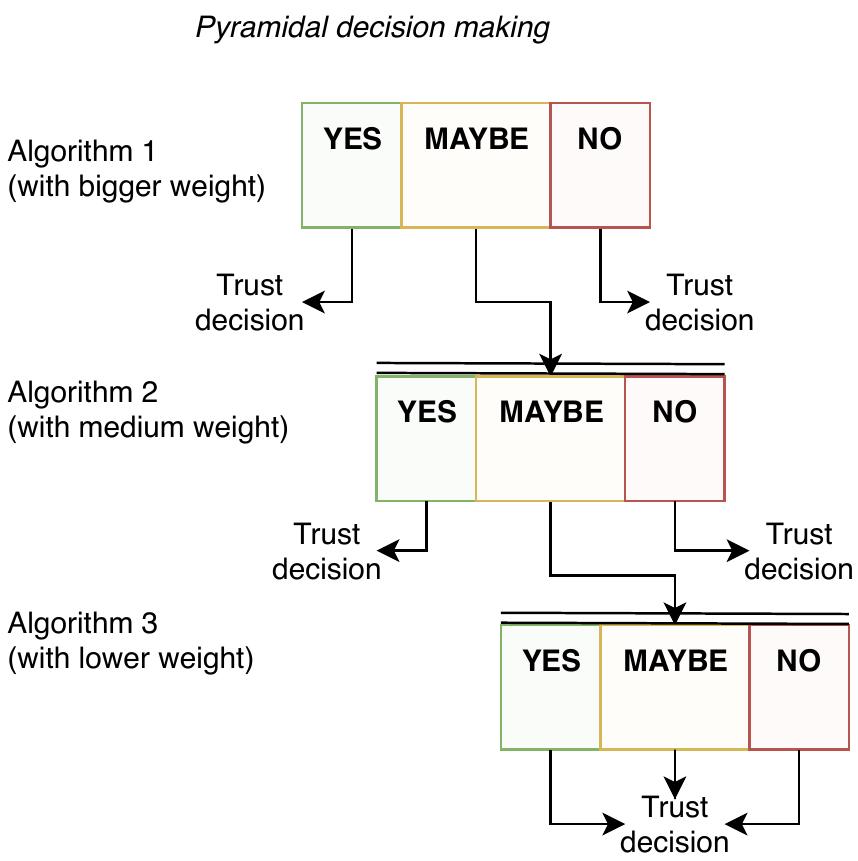} 
\caption{Pyramidal method. Please note that if more than one algorithm has a given weight, a majority method is used to get the decision "of this particular level"} \end{subfigure}
\caption{Decision making methods} 
\label{decisionamakingmethods}
\end{figure}

Decisions can be merged with distinct approaches (See Figure \ref{decisionamakingmethods}):  
\begin{itemize}[noitemsep]
\item Pareto rule (if 80\% algorithms give the same decision, we output this decision);
\item Majority rule (most prevalent decision is returned);
\item Weighted majority;
\item Pyramidal (we check for high weight algorithms, and if unsure, we check others).
\end{itemize}

\subsection{Self-Calibration}
\label{selfcalibration}

The calibration algorithm needs a set of pictures, a ground truth file (a JSON specifying the clustered version of these pictures) and some directives (Minimum true positive rate, acceptable false positive rate, minimum true negative rate, acceptable false negative rate) in order to create a configuration file with thresholds as close as possible to the expected performances.

This calibration dataset needs to be sampled from the future production dataset. Due to heavy and repeated computations, the number of sampled pictures should not exceed 100 in order for the calibration to be computed in relevant time. Exact computation time heavily depends on the enabled algorithms, hardware specifications and dataset.

Internally, the calibration algorithm will send pictures to the database, will extract the similarity graph and will evaluate this similarity graph in comparison of the ground truth reference provided. 

\begin{figure}[h!]
\centering 
\includegraphics[width=0.8\textwidth]{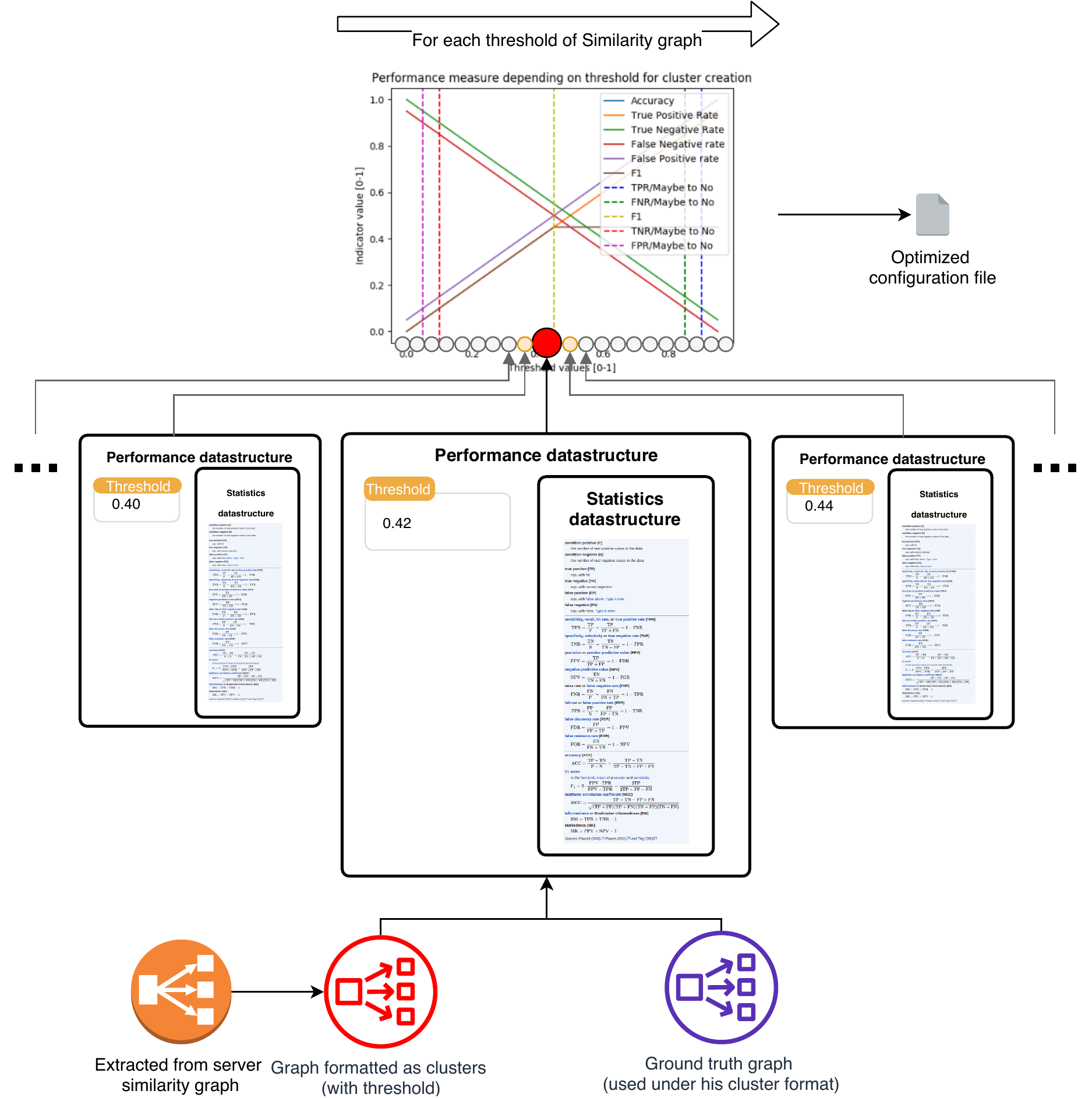} 
\caption{Calibration process overview}
\label{calibrationprocessoverview}
\end{figure}

As a comparison between a graph and a list of cluster of pictures is not trivial, we prune the graph (of distances) with a threshold. Then, we can convert the graph into clusters and compare clusters to ground truth clusters. This gives us some metrics (True Positive rate, False Positive rate, etc.) for this specific threshold. We can generate an evolution graph of these metrics. See Figure \ref{calibrationprocessoverview}. All algorithms generate the same pattern : 
\begin{itemize}[noitemsep]
\item \textbf{False positive rate} of matches increases as threshold increases. There are more and more wrong matches as we increase the acceptable distance for a match to be considered as a match.
\item \textbf{False Negative rate} of matches decreases as threshold increases. At the opposite, there are less and less matches marked as "wrong" wrongly, as we increase the acceptable distance for a match to be considered as a match.
\item \textbf{True positive rate} of matches increases as threshold increases. There are more and more matches marked as matches, which are effectively true.
\item \textbf{True Negative rate} of matches decreases as threshold increases. At the opposite, there are less and less matches that are marked as no-matches. Therefore, the number of matches marked as "no-matches" and that are truly not matches diminishes.
\end{itemize}

\begin{figure}[h!]
\centering 
\includegraphics[width=\textwidth]{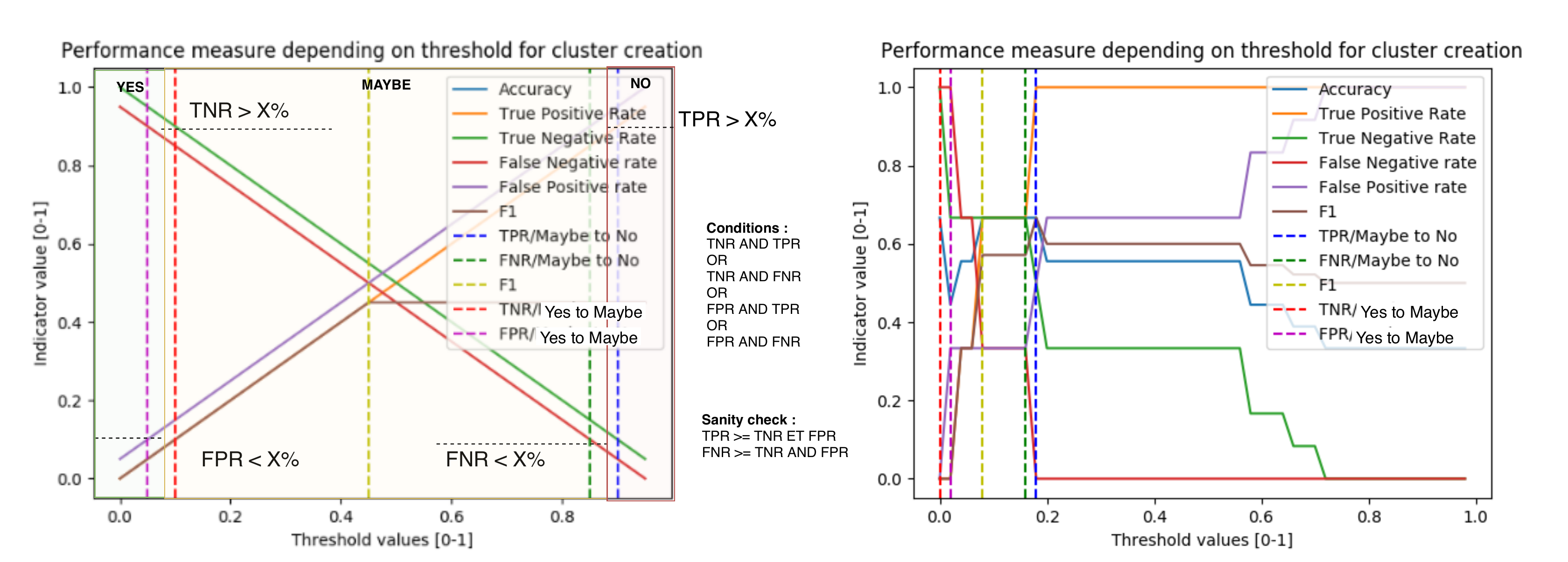} 
\caption{Thresholds extraction from a scoring graph - left values are simulated, right values are real (ORB algorithm)}
\label{optimizedmetrics}
\end{figure}

Thanks to these pattern, we can automatically find the best thresholds that optimizes this or that metric (e.g. 10\% and no more false positives, etc.) as asked by the user in the very beginning.  See Figure \ref{optimizedmetrics} which shows how thresholds are set from a True Positive rate, False Negative Rate, etc. graph.

These thresholds will then represent which decision is related to which distance for a specific algorithm. For example, a distance between two pictures below the "\textit{yes-to-maybe threshold}" (Figure \ref{optimizedmetrics}, left picture, leftmost threshold, green area) will generate a "\textit{YES}" decision. A distance between the "\textit{yes-to-maybe-threshold}" and the "\textit{maybe-to-no-threshold}" (Figure \ref{optimizedmetrics}, left picture, yellow area) will generate a "\textit{MAYBE}" decision. Finally, a distance greater than the "\textit{maybe-to-no-threshold}" will generate a "\textit{NO}" decision (Figure \ref{optimizedmetrics},left picture, red area). 

If the original graph had been generated from a configuration enabling only one algorithm, we then have extracted two thresholds for this specific algorithm. If more than one algorithm were activated, we then have two thresholds for the whole output of the library. This is not used and not relevant in our setting. So, the calibration algorithm extract only thresholds per-algorithm to build the future configuration file of the server.

\clearpage
\pagebreak
\section{Results}

Performance's related issues were encountered : 
\begin{itemize}[noitemsep]
\item \textbf{Cluster's representative picture re-evaluation is done right after the picture addition.} Therefore, the time to perform the operation adds up to adding time. This overhead could be removed from calculation by deporting the reevaluation in off-time, by adding this reevaluation job to a queue and add a new worker to complete it.
\item \textbf{Clusters number and their sizes is not deterministic.} Therefore, the time taken to perform operations is not deterministic but is bounded by $O(\textit{time to compare feature} * \textit{nb cluster} + \textit{nb good cluster} * \textit{nb pictures in each of those clusters})$
\end{itemize}

\subsection{Quality}

\begin{figure}[h!]
\centering 

\begin{subfigure}[b]{0.39\textwidth} \centering 
\includegraphics[width=\textwidth]{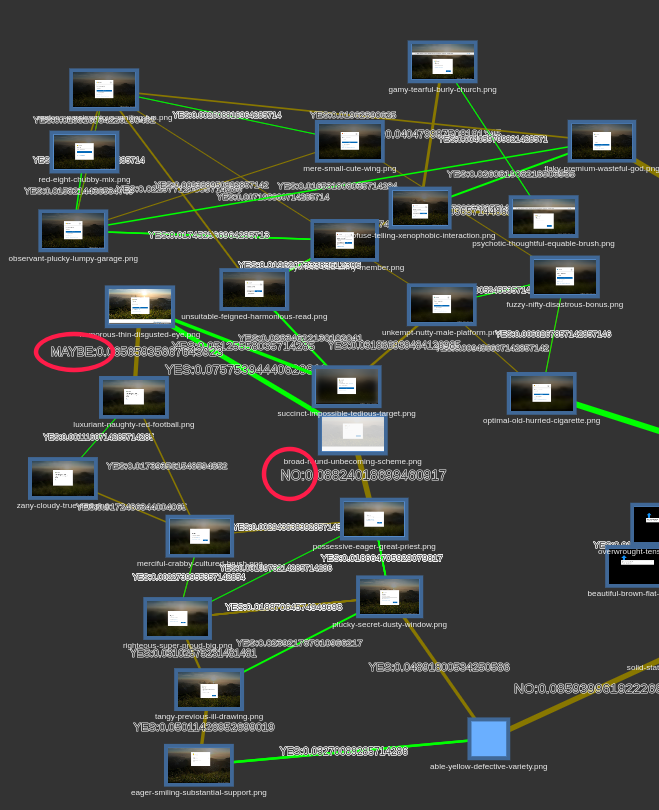} 
\caption{No problem for microsoft, even with colors, etc.} 
\end{subfigure}
\hfill
\begin{subfigure}[b]{0.59\textwidth} \centering 
\includegraphics[width=\textwidth]{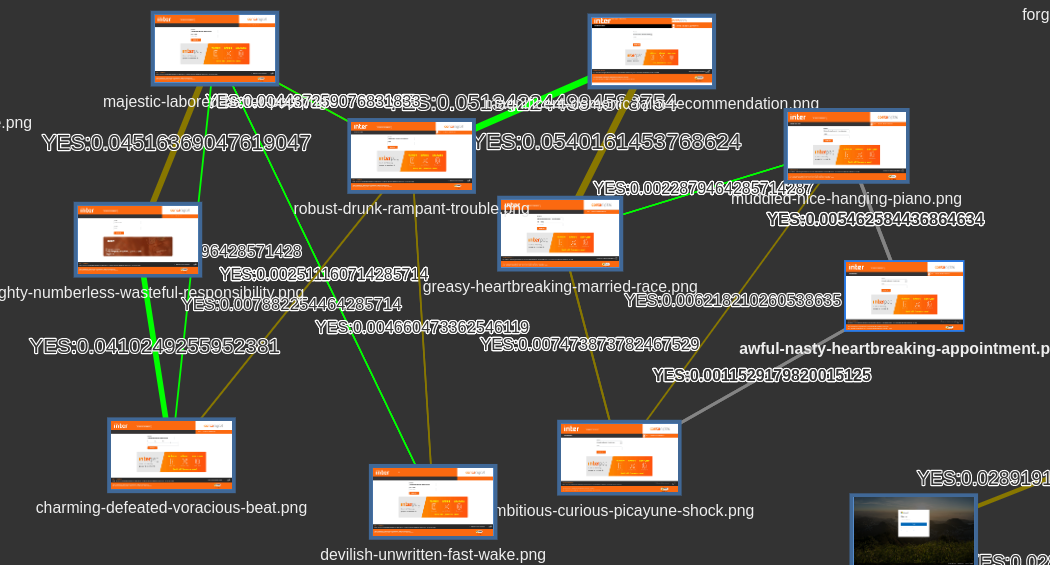} 
\caption{Good for uniform picture, even with ads} 
\end{subfigure}

\begin{subfigure}[b]{0.49\textwidth} \centering 
\includegraphics[width=\textwidth]{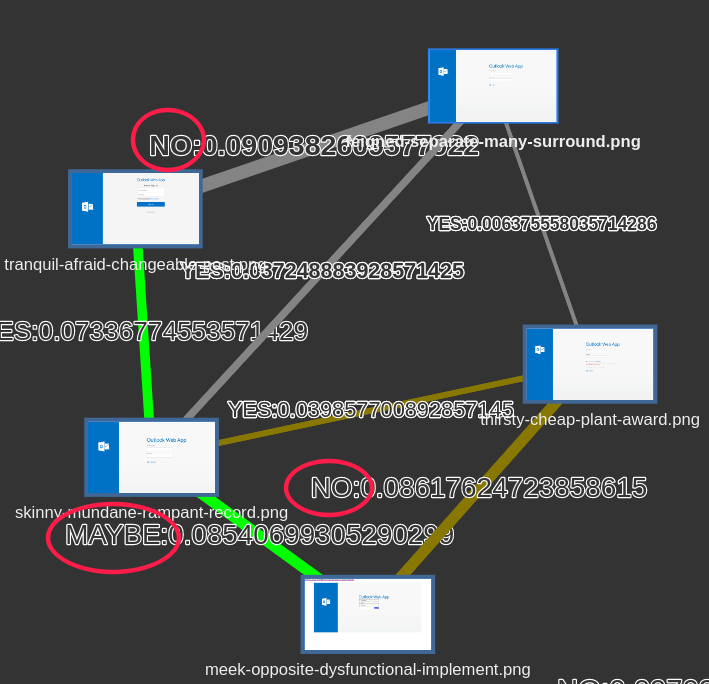} 
\caption{Good with outlook, even with modifications} 
\end{subfigure}
\hfill
\begin{subfigure}[b]{0.49\textwidth} \centering 
\includegraphics[width=\textwidth]{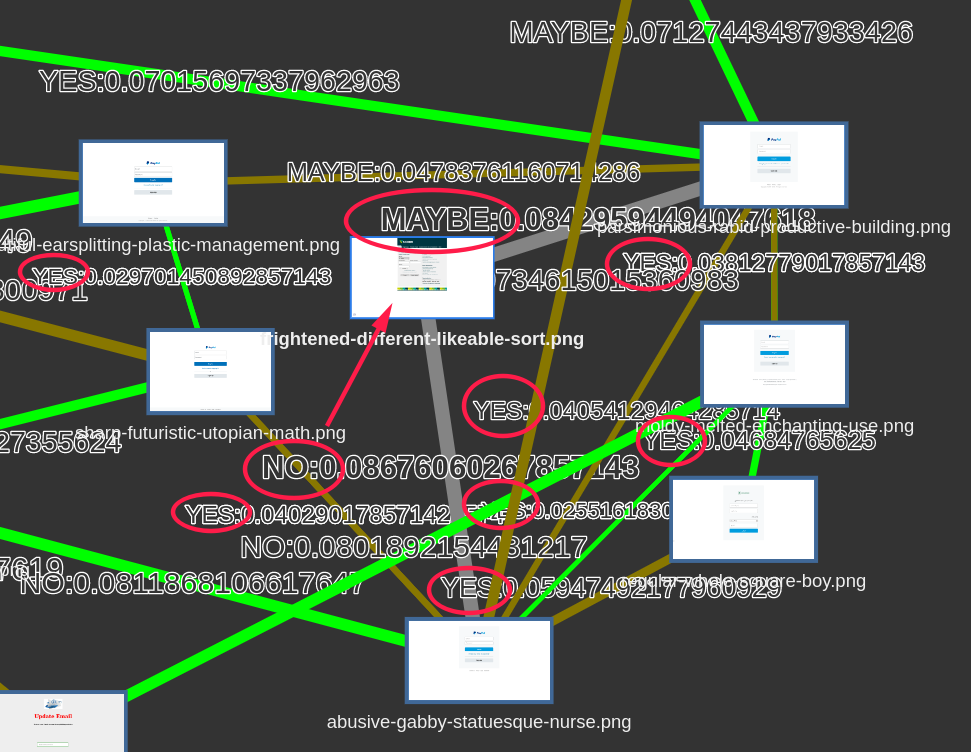} 
\caption{Positive consequence of Yes/Maybe/No: it's here possible to differentiate Paypal and non-Paypal pictures. Distance threshold would not have been enough for that.} 
\end{subfigure}

\caption{Similarity graph extracts} 
\label{setpictures1}
\end{figure}

\begin{figure}[h!]
\centering 

\begin{subfigure}[b]{0.9\textwidth} \centering 
\includegraphics[width=\textwidth]{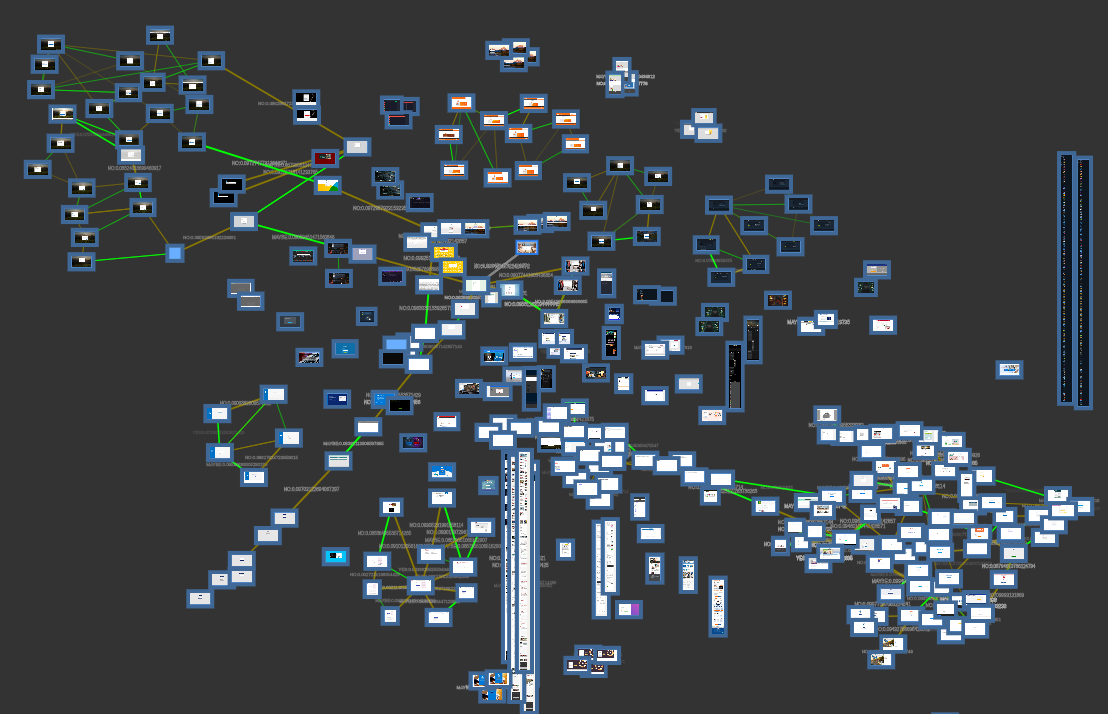} 
\caption{Global results overview - YES and MAYBE matches} 
\label{setpictures2}
\end{subfigure}
\hfill
\begin{subfigure}[b]{0.9\textwidth} \centering 
\includegraphics[width=\textwidth]{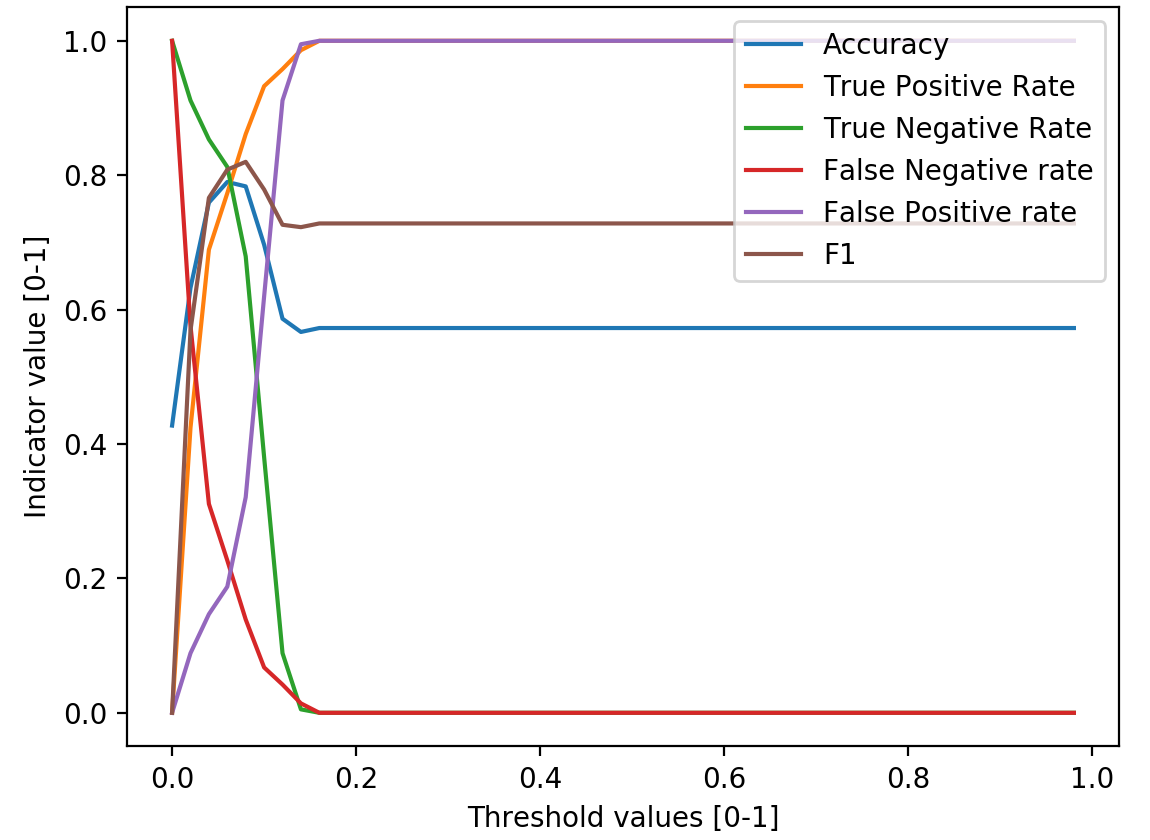} 
\caption{Matching quality regarding distance threshold} 
\label{setstats1}
\end{subfigure}

\caption{Performance overview} 
\end{figure}

\begin{figure}[h!]
\centering 

\begin{subfigure}[b]{0.49\textwidth} \centering 
\includegraphics[width=\textwidth]{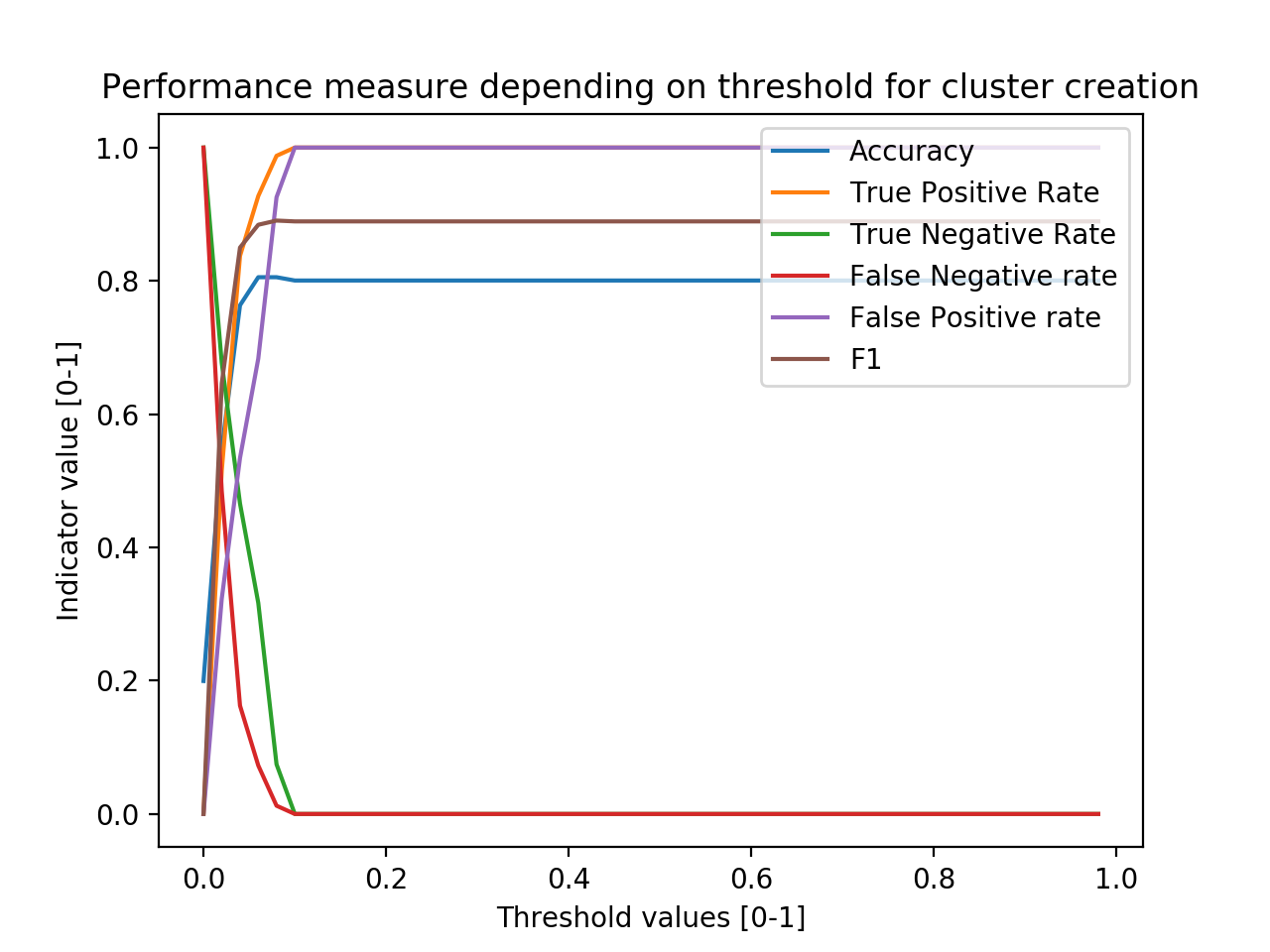} 
\caption{YES matches only} 
\label{YESONLY}
\end{subfigure}
\hfill
\begin{subfigure}[b]{0.49\textwidth} \centering 
\includegraphics[width=\textwidth]{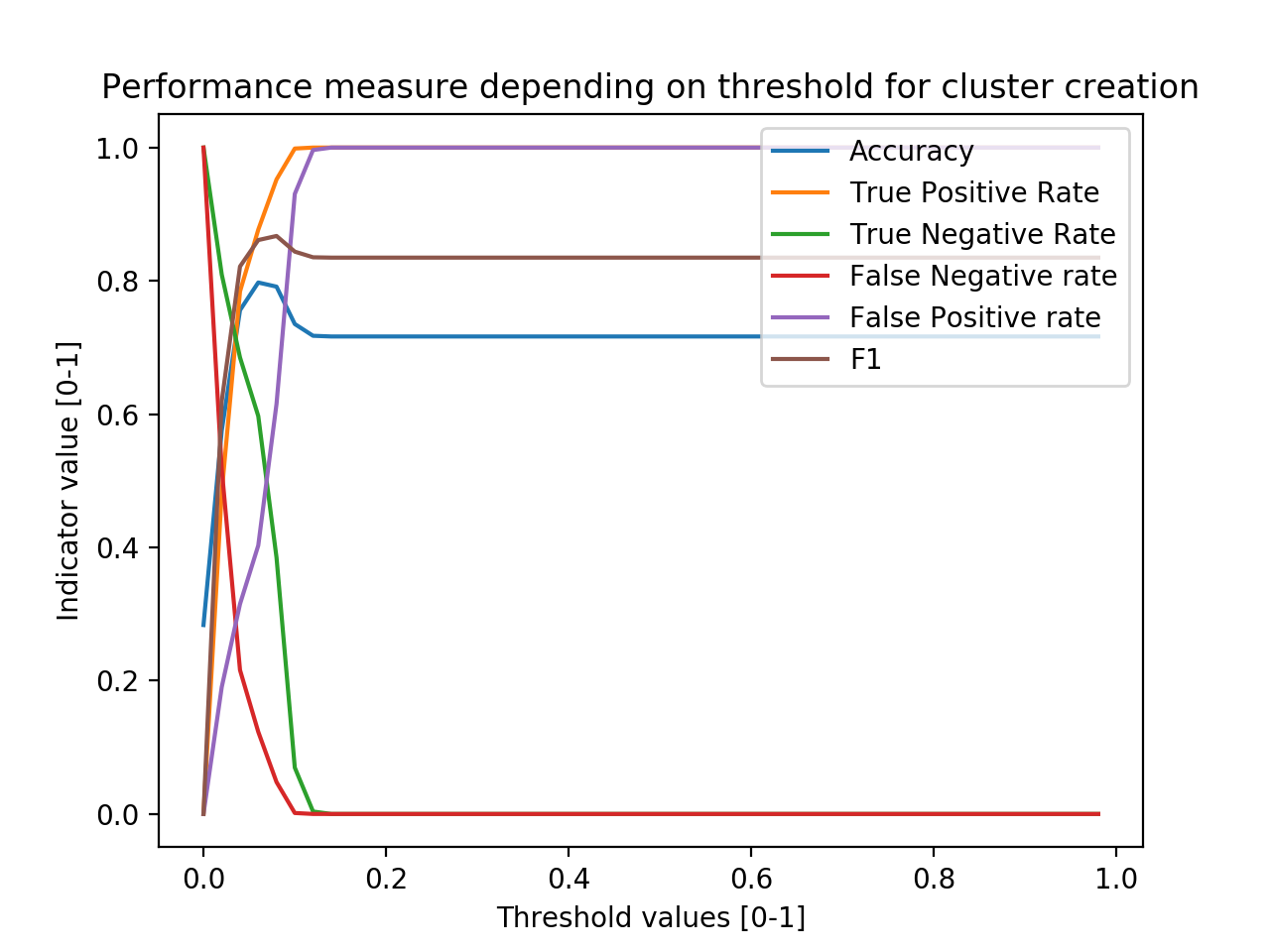} 
\caption{YES and MAYBE matches only} 
\label{YESMAYBEONLY}
\end{subfigure}

\begin{subfigure}[b]{0.49\textwidth} \centering 
\includegraphics[width=\textwidth]{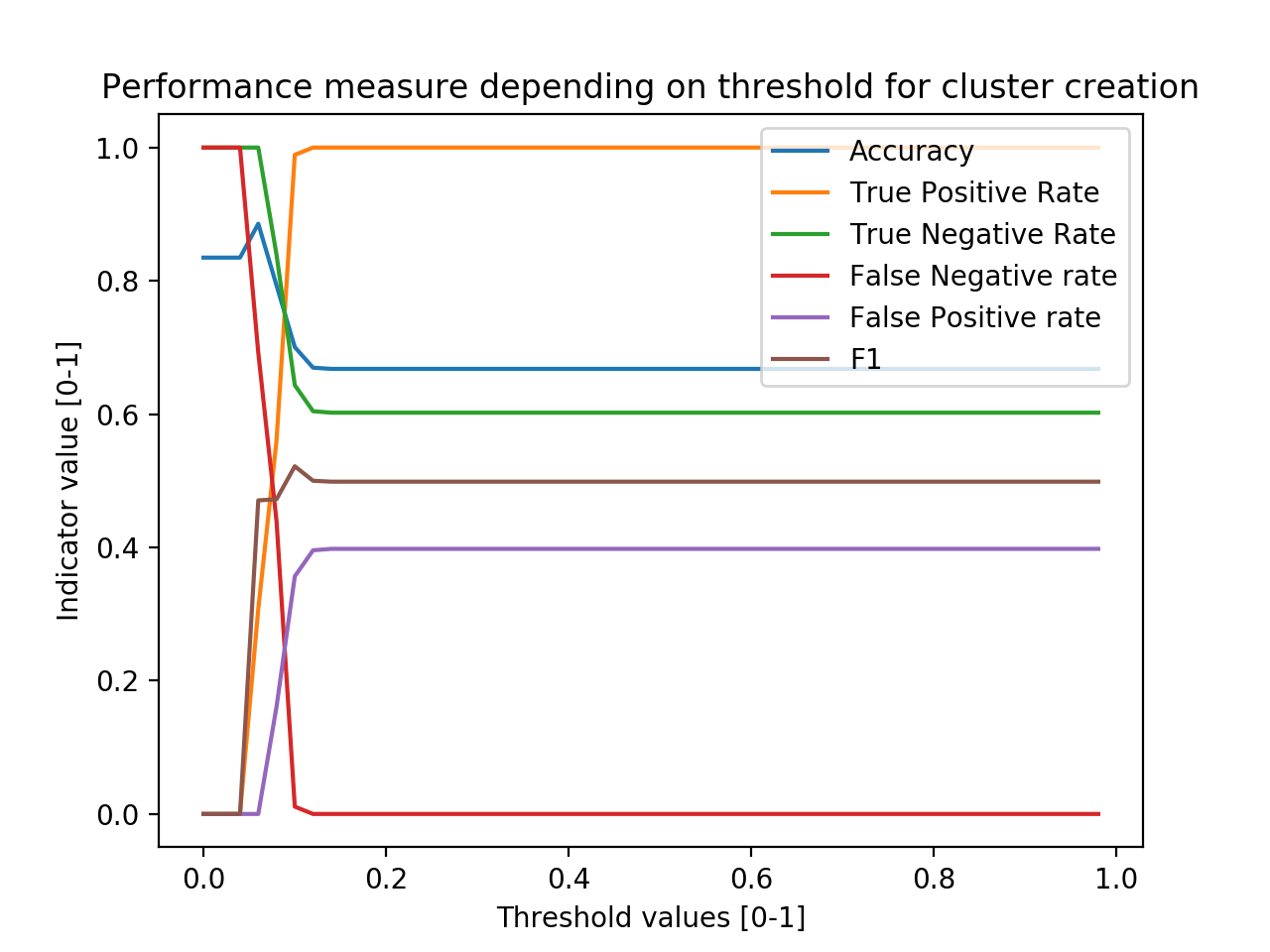} 
\caption{MAYBE matches only} 
\label{MAYBEONLY}
\end{subfigure}
\hfill
\begin{subfigure}[b]{0.49\textwidth} \centering 
\includegraphics[width=\textwidth]{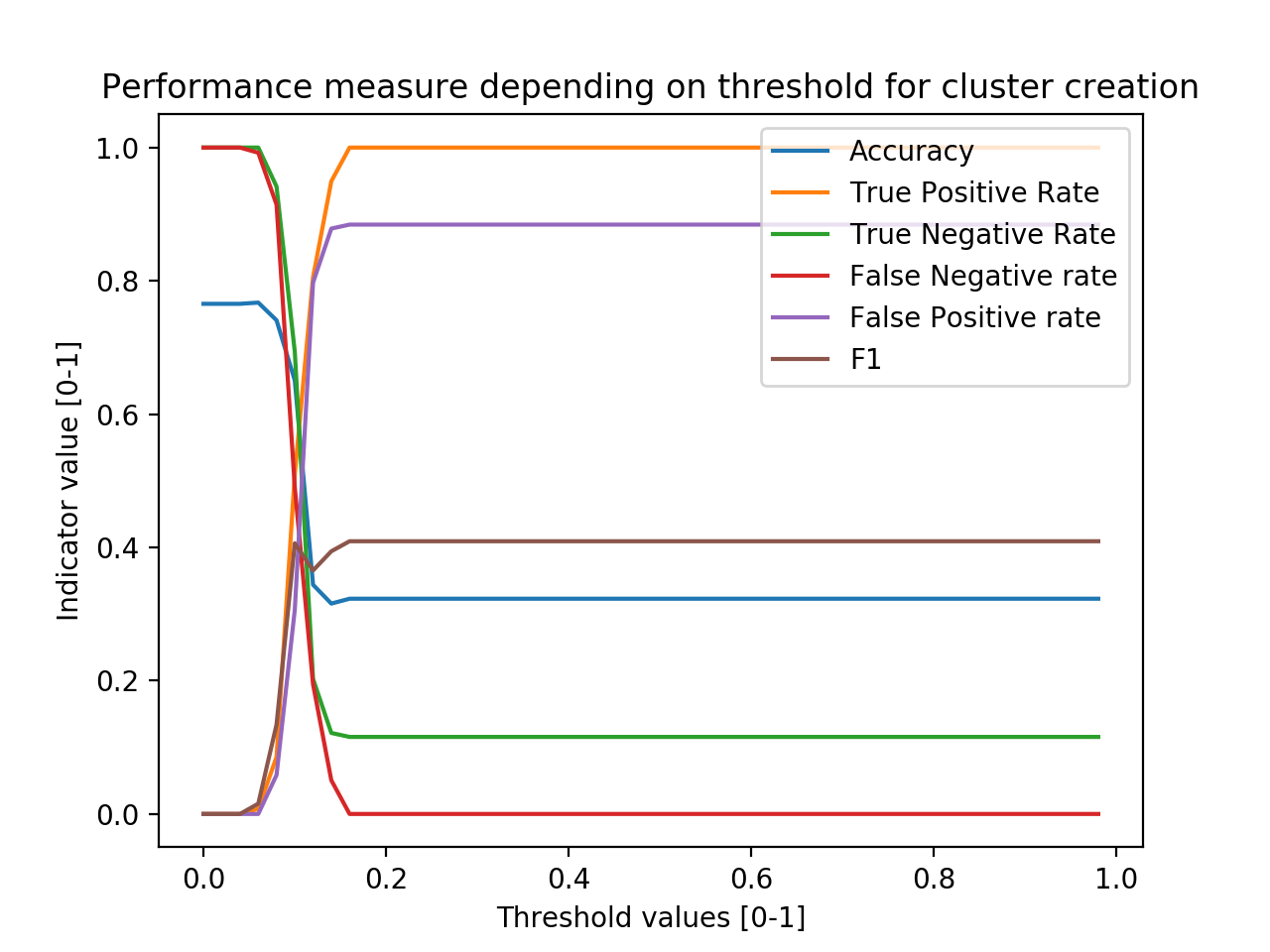} 
\caption{NO matches only} 
\label{NOONLY}
\end{subfigure}
\caption{Quality depending of threshold filtered by match type.} 
\label{setstats2}
\end{figure}

We present the performances in two ways.
\begin{itemize}[noitemsep]
\item a visualization of similarity graph with \glsenablehyper\gls{VisJS-Classificator} as shown in Figures \ref{setpictures1}, \ref{setpictures2}.
\item a quantified analysis as shown in Figures \ref{setstats1} and \ref{setstats2}.
\end{itemize}

The visualization with \glsenablehyper\gls{VisJS-Classificator} presents a graph of pictures. Each edge between two pictures represents a match. Decision and distance are written on each edge. For more information, please see \cite{falconieriVisJSClassificatorManualVisual2019}. One can easily spot issues with this view.

Figures \ref{setstats1} and \ref{setstats2} present values from a confusion matrix. 
Figure \ref{setstats1} shows that if we rely \textbf{only} on distances and put a threshold to evaluate the "goodness" of matches, we can at best reach a 80\% accuracy, 80\% F1 score with 80\% true negative, 80\% true positive, 20\% false positive and 20\% false negative (threshold at 0.075, roughly).

Figure \ref{setstats2} presents the same kind of chart for each matches subset : \textit{YES}, \textit{YES and MAYBE}, \textit{MAYBE}, and \textit{NO} matches.

\begin{itemize}[noitemsep]
\item On Figure \ref{YESONLY} (\textit{YES only}), we see that all merged distances are below 0.2. We see that if we do not put any threshold and so accept all YES matches as matches, regardless of their distances, we get a 80\% accuracy.
\item On Figure \ref{YESMAYBEONLY} (\textit{YES and MAYBE only}), we see that if we do not put any threshold and so accept all YES or MAYBE matches as matches, regardless of their distances, we get a 75\% accuracy.
\item On Figure \ref{MAYBEONLY} (\textit{MAYBE only}), we see that if we do not put any threshold and so accept all MAYBE matches as matches, regardless of their distances, we get a 65\% accuracy.
\item On Figure \ref{NOONLY} (\textit{NO only}), we see that if we do not put any threshold and so accept all NO matches as matches, regardless of their distances, we get a 35\% accuracy.
\end{itemize}

\subsection{Speed}

One of our findings is that ORB (OpenCV2 descriptors matcher) is the most time-intensive operation during adding of a picture to the database. This "\textit{Are these descriptors matching ?}" operation is called several time during the addition of a picture to the database.
Solutions to this issue are : 
\begin{itemize}[noitemsep]
\item \textbf{Reducing the time needed to match descriptors together.} As this function is part of OpenCV2 library, this is not an easy solution. Decreasing the number of descriptors per picture is a partial solution, but with a cost in picture's description's quality.
\item \textbf{Reducing the number of calls to OpenCV2 matcher.} This is intrinsically related to the database's data-structures. This would ask to modify the internal storage structure of pictures, for example for a tree-like structure or a hash-table-like structure. Due to other constraints (ability to combine algorithm outputs, etc.) this would ask for major changes in the library design.
\item \textbf{Removing ORB algorithm.} This is the most straightforward solution. However, we would loose a complementary and pertinent matching algorithm. A potential replacement for this loss would be an ORB-BoW algorithm. \glsenablehyper\gls{Carl-Hauser} showed that this version of the algorithm, using a boolean vector of presence/non-presence of a finite amount of descriptor only, is as fast as its fuzzy-hashes counterparts. 
\end{itemize}

Please note that the evaluation of speed is performed as follow : 
\begin{itemize}[noitemsep]
\item A folder of picture is divided in "\textit{boxes}" of incremental sizes, logarithmic or not. \\\textit{E.g. a first box of 10 pictures, a second of 25, a third of 40, etc.}
\item For each boxe size : 
	\begin{itemize}[noitemsep]
	\item We put aside 10 pictures that will serve as request pictures.
	\item We also put aside $N$ pictures, the exact number being the current box size.
	\item We send the $N$ pictures in the adding queue to the database, and wait until they are added.
	\item We request the 10 pictures, and wait until they have ready results.
	\end{itemize}
\end{itemize}

One issue encountered was the cache remembering which picture has already been requested, which is a side-effect of the data-structure used. Therefore, at each iteration, we need to empty the cache or we need to request new pictures. The later was chosen for its simplicity of implementation.

\clearpage
\pagebreak

\begin{figure}[h!]
\centering 
\begin{subfigure}[b]{\textwidth} \centering 
\includegraphics[width=\textwidth]{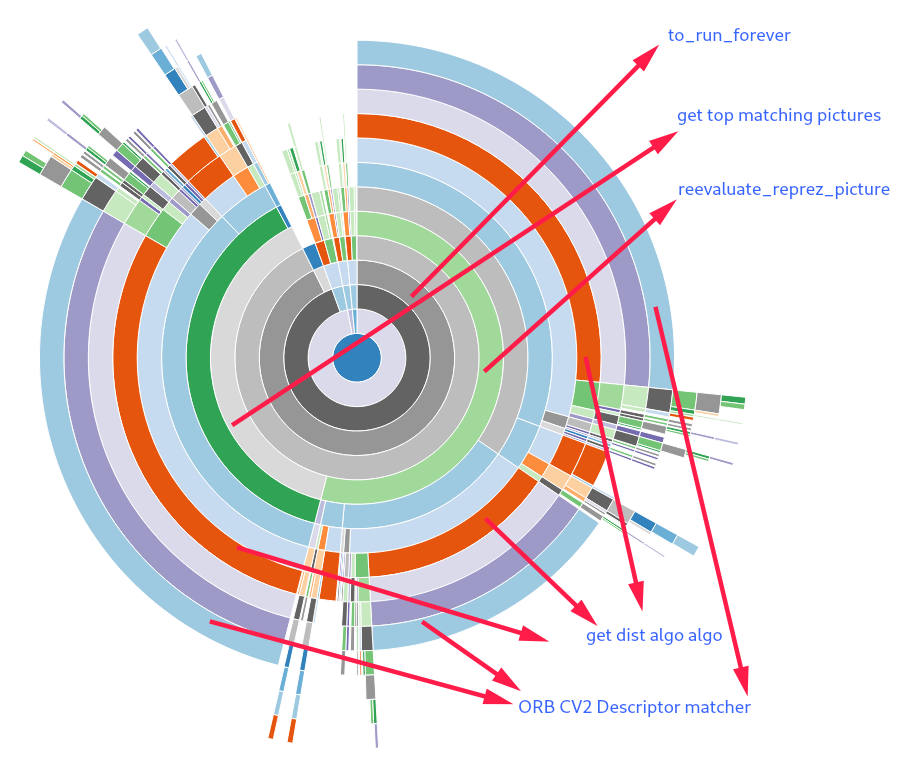} 
\caption{Time was mainly consumed by two things : re-evaluate representative picture of a cluster in O(nb pics in this cluster) complexity and the time taken by ORB evaluation directly in OpenCV2.} 
\end{subfigure}
\hfill
\caption{Cluster creation threshold at 0 (worst case) - Time repartition in database adder - CProfile of DataBase Adder Worker displayed with Sphinx} 
\end{figure}

\section{Future work}

\subsection{Potential extensions}

Some elements can be added to the current state of the \glsenablehyper\gls{Douglas-Quaid} library.
\begin{itemize}
\item \textbf{Add a "on demand" algorithm calculation mechanism}. Some algorithms need a lot (relative to others) of time to compute an output between two pictures. These algorithms (RANSAC, ORB with some modifications, etc.) may need to be triggered only for very special purposes, and very sparsely. For example, we could envision an "inverse pyramidal matching": when the most trusted algorithm outputs a "\textit{MAYBE}", it triggers the computation of one of these computation-intensive algorithms.
\item New core algorithms could be added, as : 
	\begin{itemize}
	\item \textbf{OCR and Cosinus Distance}: It would be possible to add an algorithm to store raw text as an additional feature vector, and then use cosinus distance between these set of words. Standard methods (e.g. Tesseract) could be used for the OCR, as well as State of art approach (Machine Learning with ConvNets, LTSM ...). The last ones may however be patented and can't fit into the project license requirements.
	\item \textbf{Manual matching}: A tricky algorithm could be added to the library to allow a human to perform "manual" comparisons. This very simple algorithm would only check in an aside database if a human comparison was made. If so, and if the human stored "\textit{These two pictures are similar}", then the algorithm outputs a 0-distance and a "\textit{YES}" decision (perfect match). If the human stored "\textit{These two pictures are not at all similar}", then the algorithm output a 1-distance and a "\textit{NO}" decision. Otherwise, it outputs a 0.5-distance and a "\textit{MAYBE}" decision. By giving this algorithm the highest weight and choosing a pyramidal decision process, this would bypass all other algorithms in "\textit{YES}" and "\textit{NO}" cases and put human decision above all. A HCI would be needed for the human to perform these comparisons, as well as an aside data-structure to store human's decisions.
	\item \textbf{ORB, RANSAC and geometrical verification}: As tested in \glsenablehyper\gls{Carl-Hauser}, RANSAC is a very computation-intensive but relevant algorithm. From it, we can extract the transformation matrix between two pictures (strictly, the transformation matrix to apply to the position of the key-points of the first picture in order to match the position of the key-points in the second picture). This remove most of the false-positive pictures, for a high computational cost. This would need to be performed only when doubting about the quality of the match and would need deeper modifications in the library as just "a new algorithm block". This would however be a relevant algorithm to add, as "standard algorithm" for small databases of pictures (e.g. matching a specific logo in request pictures).
	\item \textbf{ORB BoW - Bag of Word approach}: As tested in \glsenablehyper\gls{Carl-Hauser}, the Bag-Of-Word approach constitutes a good trade-off between a time-efficient matching and its relative quality compared to ORB-alone algorithm. The principle is pretty simple. We compute a k-means over the list of all keypoints and their descriptors of all pictures currently in the database. This gives us $k$ representative descriptors (each being a "mean" of a group of original descriptors). Now, instead of keeping a descriptors list (each being a real value) per picture, we only keep a boolean array, which essentially says "\textit{the $i\textit{-th of }k$ descriptor is present in this picture}". This boolean array can be processed as a hash, with Hamming distance for instance. The main costs is to create this dictionary and to maintain it. A lot of various pictures are needed to create the first dictionary. At some point, if the dataset in the production database evolves too much (if it becomes too distant with the original dataset on which the actual dictionary had been computed), the dictionary would need to be recomputed (on the actual set of pictures present on the database, larger than the initial one). The feature vector of each picture would also need to be recomputed. The complexity of such operation would be $O(\textit{nb pictures in database})$, which is not an issue. On a software engineering side, choosing \textit{when} to recompute the dictionary would however be an issue as well as the choice of the \textit{size} of the dictionary (the number of "\textit{words}" or "\textit{boolean switches}"). This issue is strongly linked to the choice of k for a k-means, and so one solution could be an affinity propagation approach.
	\item \textbf{Machine-Learning and/or BlackBox algorithm}: Machine Learning with ConvNets, LTSM ... and most recent approaches could fit in as a new algorithm block. A very simple approach with Decision Trees for decision making or/and decision merging and/or distance merging could be an easy very first approach\href{https://www.datacamp.com/community/tutorials/decision-tree-classification-python}{.}
	\item \textbf{ORB and KMeans} : As explained in \cite{chenFightingPhishingDiscriminative2009}, it seems that grouping keypoints by their position by performing k-means, and checking the similarity of these groups of keypoints between two pictures, give relevant results. However, ORB-alone is computation intensive and adding overhead computation over it may forbid its usage. Like RANSAC, this could however be relevant for small database settings and/or as a final computation.
	\item \textbf{Random algorithm} : We could use a random matching algorithm (giving a random value between 0 and 1 and a random decision) to have a reference for all algorithms. That's only informative and not a primary issue.
	\end{itemize}
	
\item \textbf{Add precomputation (text hider ... )} : As tested in \glsenablehyper\gls{Carl-Hauser}, we know that some pre-computations can improve the output quality of some algorithms (e.g. Text hider for fuzzy-hashes). We may want a way to enable or disable some specific pre- or post-computations for specific algorithms.
\item \textbf{Compare video or GIFs} : even if videos are only ordered list of pictures, we could think about finding the most relevant pictures in these videos, and then perform the matching on this set of representative pictures. This would need heavy modification to the library.
\item \textbf{Auto-tagging} : by storing arbitrary data along with picture's feature, we could give back the data stored along matched pictures with request's result. This could allow, for instance, to store tags about pictures, and so transform \glsenablehyper\gls{Douglas-Quaid} into an auto-tagging system. It could returns tags from a request picture, if added pictures were tagged. This system could be built as a wrapper around \glsenablehyper\gls{Douglas-Quaid} too.
\item \textbf{Direct integration with \glsenablehyper\gls{VisJS-Classificator}} : Visualization of the links between pictures could be presented with a graph view. Calls to \glsenablehyper\gls{VisJS-Classificator} API could be performed when some key computations are performed in \glsenablehyper\gls{Douglas-Quaid} to trigger graph modification.
\item \textbf{Metrics to evaluate the state of the database} : the quality of the matching, the performances of the request system, the quality of the clustering inside the main data-structure, etc. could be evaluated at any time.
\item \textbf{Handling duplicates} : Duplicated pictures (same cryptographic hash) are already structurally handled by the library. If two same pictures are requested or added, the will have the same ID and so be "already computed". However, client is not alerted about the previous presence of this specific picture.
\item \textbf{Dynamic threshold modification} : Internal threshold of the library are set once and for all at the library first launch. Any post modification would probably generate artifacts in matching. We recommend to flush the server and add again all stored pictures if thresholds are modified. However, one may want to "\textit{do it live}". A trade off is that original pictures are needed to perform such operation. Therefore, "\textit{doing it live}" would need to have an accessible copy of all pictures ever stored in database until re-computation time, which is only optional for now.
\item \textbf{Dynamic features modification} : The problem is similar for enabled features. 
A recomputation of all features of pictures with missing algorithms is needed when we change which algorithms are enabled. So, we would need the original files.
\end{itemize}

\subsection{Planned extensions}
Problems presented previously lead to a list of future possible developments : 
\begin{itemize}
\item Extending provided datasets to support research effort
\item Improving scalability of the library
\end{itemize}

\clearpage
\section{Conclusion}
\subsection{Summary}

Our Image-Matching library named \glsenablehyper\gls{Douglas-Quaid} is available at \href{https://www.github.com/CIRCL/douglas-quaid}{github.com/CIRCL/douglas-quaid}. This research paper proposes that even partial automation of screenshots classification would reduce the burden on security teams and that \glsenablehyper\gls{Douglas-Quaid} is a step forward in this direction.

The original algorithms evaluation framework named \glsenablehyper\gls{Carl-Hauser} is available at \href{https://www.github.com/CIRCL/carl-hauser}{github.com/CIRCL/carl-hauser}.

The datasets used to conduct tests are available at : 
\href{https://www.circl.lu/opendata/circl-ail-dataset-01/}{circl.lu/opendata/circl-ail-dataset-01} and \href{https://www.circl.lu/opendata/circl-phishing-dataset-01/}{circl.lu/opendata/circl-phishing-dataset-01}

Research results were presented along with encountered issues. Algorithmic modifications are still being performed to solve met issues and improve overall quality of the library. This library is intended to be used in Open Source tools or for research purposes.

\subsection{Contact information}
\label{contactinfo}
If you have a complains related to the dataset or the processing over it, please contact us. We aim to be transparent, not only about how we process but also about rights that are linked to such information and processing. 
You can contact us at \href{https://circl.lu/contact/}{circl.lu/contact/} for request about the dataset itself, regarding elements of the dataset, or extension requests.
You can contact us at same address or on \href{https://github.com/CIRCL/douglas-quaid}{github} for feedback about the benchmarking framework, methodology or relevant ideas/inquiries.

\clearpage
\pagebreak

\bibliographystyle{paper-ressources/IEEEtran}
\bibliography{./carl-hauser.bib}

\clearpage
\pagebreak
\part{Appendices}

\section{Detailed view of the Library}
\label{app:view_framework}

The exact implementation differs on some points with the idealized overview presented in the paper's body. 

\begin{sidewaysfigure}[h!]
\centering 
\includegraphics[width=\textwidth]{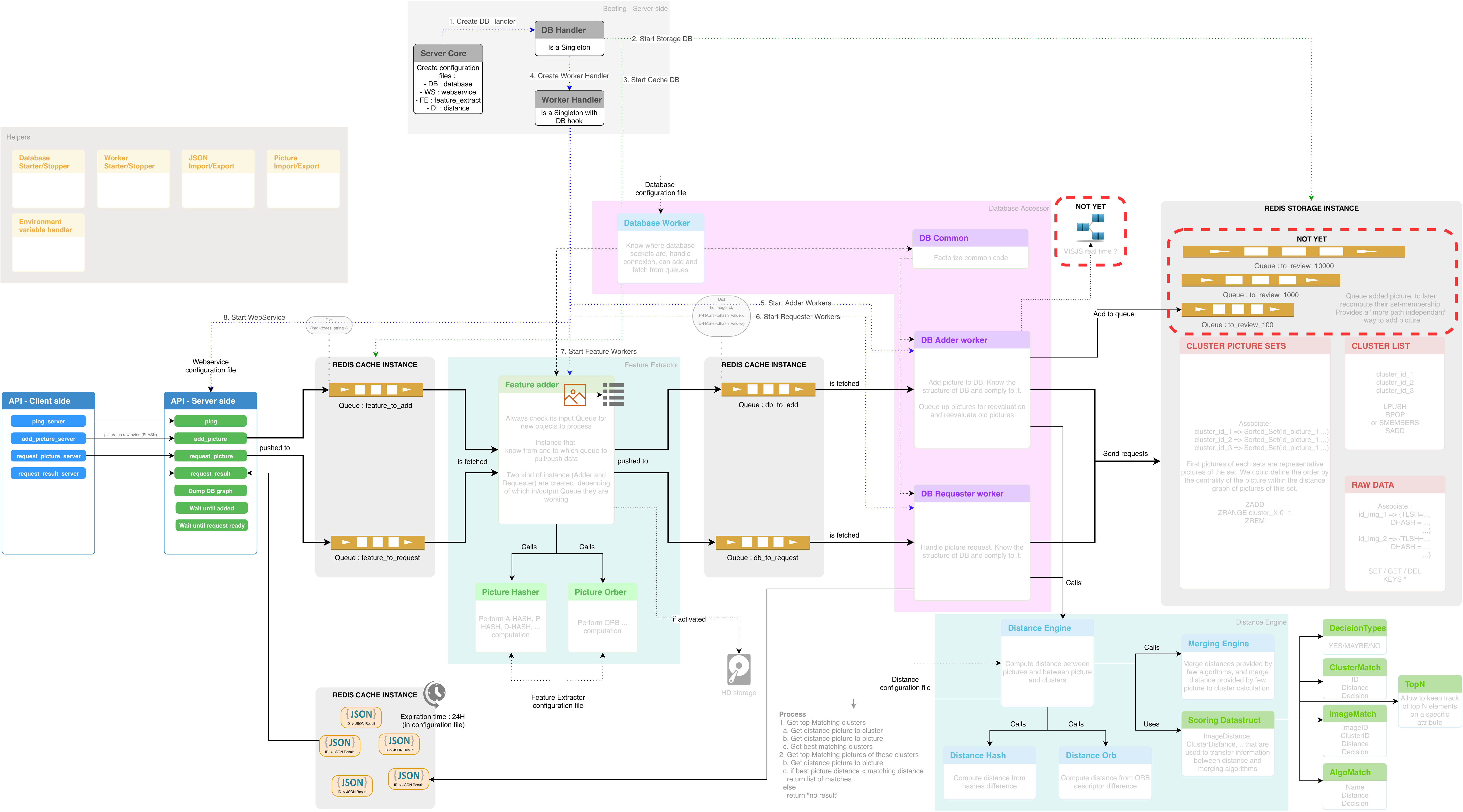} 
\caption{Software Components architecture} 
\label{overviewLib}
\end{sidewaysfigure}

For example, the queuing system is more complex than just one queue of jobs. It is divided in many smaller queues for each kind of jobs. Workers are of different types depending on the process step in the pipeline. Other mechanisms are presents, as refreshing and re-evaluation algorithms to prevent path-(upload)-dependent behaviors.
Algorithms-specific classes are modular and aside from the rest of the application. Adding an algorithm (OCR, human-matching, ML, ...) would only take two files creations, ask for modification of two more files and less than 100 lines of "context management". See Figure \ref{overviewLib} for the detailed view of the library processes.

\begin{sidewaysfigure}[h!]
\centering 
\includegraphics[width=1\textwidth]{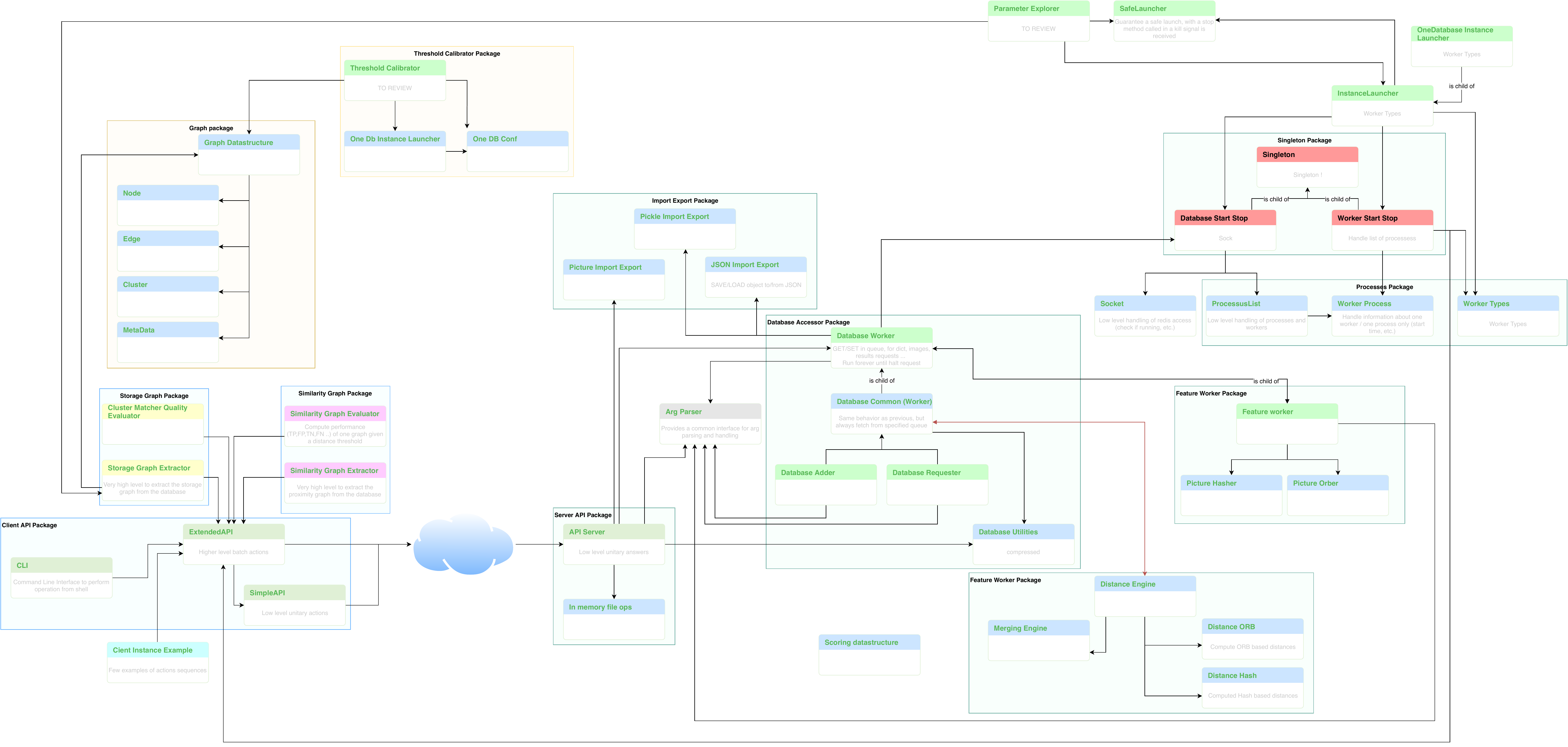} 
\caption{CLI, API and extractor structure on client side}
\label{model}
\end{sidewaysfigure}

The library classes model is shown in Figure \ref{model} and presents which classes is using which other classes.

\section{Client API Usage example}

\lstset{language=C++,
                basicstyle=\ttfamily,
                keywordstyle=\color{blue}\ttfamily,
                stringstyle=\color{red}\ttfamily,
                commentstyle=\color{green}\ttfamily,
                morecomment=[l][\color{magenta}]{\#}
}

\begin{lstlisting}[frame=single,caption=Code example of a client side client,label=exampleclientapi, basicstyle=\footnotesize]
    @staticmethod
    def example():
        # Generate the API access point link to the hardcoded server
        cert = (get_homedir() / "carlhauser_client" / "cert.pem").resolve()
        api = Simple_API(url='https://localhost:5000/', certificate_path=cert)

        # Ping server, and perform uploads
        api.ping_server()
        api.add_one_picture(get_homedir() / "image.jpg")
        # (...)

        # Request a picture matches
        request_id = api.request_similar(get_homedir() / "image.bmp")[1]
        # (...)

        # Wait a bit
        api.poll_until_result_ready(request_id, max_time=60)

        # Retrieve results of the previous request
        api.get_results(request_id)

        # Triggers a DB export of the server as-is, to be displayed with
        # visjsclassificator. Server-side only operation.
        api.export_db_server()
\end{lstlisting}

\pagebreak
\part{Glossary}
\printglossary[title=]

\end{document}

%% file: glossaryEN.tex
\newglossaryentry{TLP}
{
  name={TLP},
  description={The Traffic Light Protocol (TLP) was created to facilitate information sharing. TLP is a set of designations used to ensure that sensitive information is shared with the appropriate audience. It uses four colours to indicate the expected sharing limits to be applied by the beneficiary(ies). TLP has only four colours; all designations not listed in this standard are not considered valid by FIRST. Translated from \cite{TrafficLightProtocol}}
}
\newglossaryentry{Sand-box}
{
  name={Sand-box},
  description={A Sand-Box is a security mechanism to separate programs running, usually to mitigate system failures or software vulnerabilities. It is often used to run untested or unreliable programs or code, possibly from third parties, suppliers, users or websites that are not verified or at risk, without risking harm to the host machine or operating system. Translated from \cite{SandboxComputerSecurity2019}}
}
\newglossaryentry{PFE}
{
  name={PFE},
  description={Research or research and development work, carried out individually or in very small groups over 4 to 6 months of full-time equivalent. The IEPTs take place in a laboratory of INSA Lyon or in a company. In all cases, they are supervised and supervised by INSA Lyon teacher-researchers and give rise to a report and defence. From \cite{ProjetsFinEtude2009}}
}
\newglossaryentry{TLSH}
{
  name={TLSH},
  description={TLSH is a fuzzy-hashing library. From a bytes stream with a minimum length of 50 bytes, TLSH generates a hash value that can be used for similarity comparisons. Similar objects will have similar hash values, which allows similar objects to be detected by comparing their hash values. Translated from \cite{TrendmicroTlsha}}
}
\newglossaryentry{SSDeep}
{
  name={SSDeep},
  description={SSDeep is a hash calculation program of the type "context triggered piecewise hashes" (CTPH). Also called fuzzy hashing, a CTPH algorithm can identify similar entries. These entries are sequences of identical bytes in the same order, although the bytes between these sequences may be different in content and length. Translated from \cite{SsdeepFuzzyHashing}}
}
\newglossaryentry{SIFT}
{
  name={SIFT},
  description={Method of extracting distinctive invariant elements from images that can be used to make a reliable comparison between different views of an object or scene. The characteristics are invariant to the scale and rotation of the image, and have been shown to provide a robust match over a range of sharp distortion, 3D modification, noise addition and lighting modification. Translated from \cite{ScaleinvariantFeatureTransform}}
}
\newglossaryentry{DigInPix}
{
  name={DigInPix},
  description={Automatic system capable of identifying visual entities appearing in images and videos, from a list of 25,000 entities, aggregated from Wikipedia, and more specific websites. DigInPix is a generic application designed to identify different types of entities, designed by and for the INA. Translated from the \cite{DigInInPixVisualNamedEntities2015}}
}
\newglossaryentry{SYVICOL}
{
  name={SYVICOL},
  description={Syndicat des Villes et Communes du Luxembourg \cite{AccueilSyvicol}}
}
\newglossaryentry{SIGI}
{
  name={SIGI},
  description={Intercommunal Computer Management Union \cite{Intercommunal Management Union}}
}

\newglossaryentry{Image-Retrieval}
{
  name={Image-Retrieval},
  description={An image search system is one of the solutions that can be used to navigate, search and retrieve images from a large database of digital images. Translated from \cite{marshallSurveyImageRetrieval}}
}
\newglossaryentry{Image-Matching}
{
  name={Image-Matching},
  description={Image matching is a classic problem in scene analysis: if you give a reference image of an object, you have to decide if it exists in a scene image being analyzed, and find its location if it does. The measurement of similarity between two images is an objective of Image-Matching. Translated from \cite{douRobustImageMatching2018}}
}
\newglossaryentry{Image-Classification}
{
  name={Image-Classification},
  description={Image classification is a process of grouping pixels into several classes based on the application of logical decision rules. Translated from \cite{ImageClassificationOverview}}
}
\newglossaryentry{Open-Set-Classification}
{
  name={Open-Set-Classification},
  description={For some classification problems, we do not know all the possible classes during the labeling phase. An Open-Set-Classificaiton situation is a situation where many labels are known, but more and more other labels are not yet known. Translated from \cite{scheirerOpenSetRecognition2013a}}
}
\newglossaryentry{Redis}
{
  name={Redis},
  description={Redis is an open-source in-memory database (under BSD license), used as a database, cache and message broker. It supports data structures such as strings, hashes, lists, sets, sets, sets sorted with queries by range, bitmaps, geospatial indexes with queries by radius. Redis integrates replication, Lua scripting, transactions and different levels of disk persistence, and offers high availability via Redis Sentinel and automatic partitioning with Redis Cluster. Translated from \cite{Redis}}
}
\newglossaryentry{SMILE}
{
  name={SMILE},
  description={Securitymadein.lu is the main online source for cybersecurity in Luxembourg. Its purpose is to provide news, relevant information and a toolbox with cybersecurity solutions useful for private users, organizations and the community. Translated from \cite{SECURITYMADEINLUa}}
}
\newglossaryentry{CIRCL}
{
  name={CIRCL},
  description={Computer Incident Response Center Luxembourg is a government initiative designed to provide a systematic response to IT security threats and incidents. Translated from \cite{CIRCLMissionStatement}}
}
\newglossaryentry{C3}
{
  name={C3},
  description={CyberSecurity Competence Center offers services in collaboration with partners, promotes the long-term development of wealth and skills while increasing Luxembourg's attractiveness and reputation in the field of cybersecurity. Three main objectives: observe, train and test. Translated from \cite{CybersecurityCompetenceCenter}}
}
\newglossaryentry{CASES}
{
  name={CASES},
  description={CyberSecurity Awareness and Security Enhancement Services is an initiative of the Luxembourg government that provides awareness activities, web resources and other tools such as MONARC, focusing on understanding the information security issues facing SMEs. Translated from \cite{CASESLU}}
}
\newglossaryentry{Forensic}
{
  name={Forensic},
  description={Forensics studies are a branch of digital forensics that focuses on the evidence found in computers and digital storage media. The purpose of forensics is to examine digital media in a legal manner in order to identify, preserve, retrieve, analyze and present facts and opinions about digital information. Translated from \cite{ComputerForensicsWikipedia}}
}

\newglossaryentry{RGPD}
{
  name={RGPD},
  description={The General Data Protection Regulation is a regulation of Community law on data protection and privacy for all citizens of the European Union and the European Economic Area. It also deals with the export of personal data outside the EU and the EEA. Translated from \cite{GeneralDataProtection2019}}
}
\newglossaryentry{ISO 27XXX}
{
  name={ISO 27XXX},
  description={The ISO/IEC 27000 family of standards helps organizations secure their information resources. The purpose of using this family of standards is to standardize the requirements of an information security management system (ISMS). ISO 27005 is risk management oriented. Translated from \cite{ISOIEC27001}}
}
\newglossaryentry{MISP}
{
  name={MISP},
  description={MISP, for Malware (or Magic) Information Sharing Platform, is as its name suggests a platform for sharing security events (broader than just malware).
MISP is an open source software solution that enables the collection, storage, distribution and sharing of cybersecurity indicators and threats related to the analysis of cybersecurity incidents and malware. MISP is designed by and for incident analysts, security and ICT professionals or malware reversers to support their daily operations and effectively share structured information. The objective of the MISP is to promote the sharing of structured information within the IT security community. MISP provides functionalities to support the exchange of information but also the consumption of this information by Network Intrusion Detection Systems (NIDS), LIDS but also log analysis tools, SIEMs. Translated from \cite{MISPCoreSoftware2019}\cite{wagnerMISPDesignImplementation2016}}
}
\newglossaryentry{AIL}
{
  name={AIL},
  description={AIL, for Analysis Information Leak framework, is a deep-web crawler, which will analyze the loaded content of the pages to detect possible information leaks. AIL is a modular environment for analyzing potential information leaks from unstructured data sources such as Pastebin texts, similar services or unstructured data flows. The LIA framework is flexible and can be extended to support other functionalities to extract or process sensitive information (e.g. data leak prevention). \cite{CIRCLAILAnalysis} for reference, translated from \cite{AILFrameworkAnalysis2019a}\cite{mokaddemAILDesignImplementation2018}}
}
\newglossaryentry{D4}
{
  name={D4},
  description={D4 is an IoT protocol analyzer that can be easily extended to any parser and/or anaylator. It is a network of widely distributed sensors to monitor DDoS and other malicious activities while remaining an open and collaborative project. D4 is therefore a set of software components, which includes everything you need to create your own sensor network or connect to an existing sensor network using simple clients. From \cite{HomeD4Project} and translated from \cite{D4projectD4coreD4}.}
}
\newglossaryentry{Lookyloo}
{
  name={Lookyloo},
  description={Lookyloo is a web site and content analyzer that is dynamically loaded. Lookyloo is therefore a web interface that allows you to crawl a website and display a tree structure of the domains that are called. Translated from \cite{LookylooWebInterface2019}}
}
\newglossaryentry{URLAbuse}
{
  name={URLAbuse},
  description={URLAbuse is a static website verification tool, which can give an indicative opinion on the probability that the evaluated site is a phishing site. URL Abuse is a free and versatile software for URL verification, analysis and blacklist creation. URLAbuse consists of a web interface where requests are submitted asynchronously and a back-end system to process URLs in a modular way. Translated from \cite{URLAbuseVersatile2019}}
}
\newglossaryentry{Monarc}
{
  name={Monarc},
  description={Developed by \glsenablehyper\gls{CASES}, Monarc \cite{g.i.eMONARC} is a risk assessment and management tool, in conjunction with ISO certifications, standards (including GDPR), etc. The software allows a complete risk analysis to be carried out on the company. The advantage of MONARC is the capitalisation of risk analyses already carried out in similar business contexts: the same vulnerabilities regularly appear in many companies, as they face the same threats and generate similar risks. Most companies have servers, printers, a fleet of smartphones, Wi-Fi antennas, etc. so the vulnerabilities and threats are the same. It is therefore sufficient to generalize the risk scenarios for these assets (also called objects) by context and/or activity. Translated from \cite{MONARCMethodOptimised2019}}
}
\newglossaryentry{MISP-Modules}
{
  name={MISP-Modules},
  description={MISP modules are stand-alone modules that can be used for the expansion and provision of other services. The modules are written in Python 3 following a simple API interface. The objective is to facilitate the extension of MISP functionalities without modifying the basic components. The API is available via a simple REST API that is independent of the installation or MISP configuration. Translated from \cite{ModulesExpansionServices2019}}
}
\newglossaryentry{MISP-Taxonomies}
{
  name={MISP-Taxonomies},
  description={MISP Taxonomies is a set of common classification libraries for tagging, classifying and organizing information. Taxonomy allows the same vocabulary to be expressed among a distributed set of users and organizations. These taxonomies can be used in MISP and other information sharing tools and expressed in labels (triplets). A machine tag is composed of a namespace (MUST), a predicate (MUST) and a value (OPTIONAL). Machine tags are often called triple tags because of their format. Translated from \cite{TaxonomiesUsedMISP2019b}}
}

\newglossaryentry{Carl-Hauser}
{
  name={Carl-Hauser},
  description={Open-Source Framework for testing algorithms for correlation, distance and image analysis. Strongly linked to: Douglas-Quaid. Translated from \cite{OpenSourceTesting2019}}
}
\newglossaryentry{Douglas-Quaid}
{
  name={Douglas-Quaid},
  description={Free software for correlation, distance and image analysis. Strongly linked to: Carl-Hauser. Translated from \cite{OpenSourceSoftware2019a}}
}
\newglossaryentry{VisJS-Classificator}
{
  name={VisJS-Classificator},
  description={Classifier for image matching and grouping. Fast and visual. Uses VISjs and Socket.io to allow the classification/matching of images, manually, in real time in multi-user mode. translated from \cite{vincent-circlClassificatorPicturesMatching2019}}
}
\newglossaryentry{VisJS-Network}
{
  name={VisJS-Network},
  description={Tool for viewing networks composed of nodes and links. The visualization supports different shapes, styles, colors, formats, images, etc. Network visualization works without problems on any modern browser up to a few thousand nodes and edges. To manage a larger number of nodes, the network has clustering support. The network uses HTML canvas for rendering. The library is easy to modify and already used in MISP in a modified and improved version. Translated from \cite{VisJsNetworka}}
}
\newglossaryentry{Fuzzy-Hash}
{
  name={Fuzzy-Hash},
  description={Signature/hash technique that combines a number of traditional hashes whose limits are determined by the context of the entry. These signatures can be used to identify modified versions of known files even if data has been inserted, modified or deleted in the new files. This method has applications in forensic computer science. Translated from \cite{kornblumIdentifyingAlmostIdentical2006a}}
}
\newglossaryentry{ORB}
{
  name={ORB},
  description={Very fast approach based on binary descriptors based on BRIEF, which is rotationally invariant and resistant to noise. ORB is two orders of magnitude faster than SIFT, while also performing well in many situations. Can be applied to object detection or patch-tracking on a smartphone. translated from \cite{rubleeORBEfficientAlternative2011a}}
}
\newglossaryentry{Feature-Point}
{
  name={Feature-Point},
  description={A Feature-Point is defined as an "interesting" part of an image, and the characteristics are used as a starting point for many Computer Vision algorithms. Since the characteristics are used as the starting point and main primitives for the following algorithms, the processing algorithm will often be as good as its feature-point detector. Translated from \cite{FeatureDetectionComputer2019}}
}
\newglossaryentry{Hamming}
{
  name={Hamming},
  description={Hamming distance is used in telecommunication to count the number of altered bits between two messages of a given length. The distance is compatibilized as the number of differences between the two bit strings. From \cite{DistanceHamming2019}}
}

\newglossaryentry{OpenSource}
{
  name={OpenSource},
  description={Open source is a term indicating that a product includes permission to use its source code, design documents or content. It most often refers to the open-source model, in which software or other products are published under open-source license. Translated from \cite{OpenSourceWikipedia}}
}
\newglossaryentry{OpenData}
{
  name={OpenData},
  description={Open Data is the publication of data that can be used and republished without restrictions of copyright, patents or other control mechanisms. The OpenData initiative is supported by the launch of government initiatives. This data can support the research effort in particular areas by providing an essential raw material for researchers. Translated from \cite{OpenData2019}}
}
\newglossaryentry{Zotero}
{
  name={Zotero},
  description={Zotero is a free tool to help collect, organize, cite and share research sources, such as papers, theses, etc. Translated from \cite{ZoteroYourPersonal}}
}
\newglossaryentry{DataTurks}
{
  name={DataTurks},
  description={Data annotation tool for Machine Learning. The tool has been designed to allow teamwork. After sending the data to the instance and adding team members, it is possible to annotate and create a set of learning or evaluation data in a few hours.
  \cite{dataturksMLDataAnnotations2019} and translated from \cite{dataturksMLDataAnnotations2019a} }
}